\documentclass[usenatbib]{mn2e}
    \pdfoutput=1
    \usepackage{amsmath}
    \usepackage{amssymb}
    \usepackage{graphicx}

    \graphicspath{{./figures/}}

    \newcommand{\MSun}{\ensuremath{\mathrm{M}_\odot}}
    \newcommand{\RSun}{\ensuremath{\mathrm{R}_\odot}}
    
    \newcommand{\orb}{\ensuremath{_{\mathrm{orb,p}}}}
    \newcommand{\rp}{\ensuremath{r_{\mathrm{p}}}}
    \newcommand{\rpdot}{\ensuremath{\dot{r}_{\mathrm{p}}}}
    \newcommand{\zams}{\ensuremath{_{\mathrm{ZAMS}}}}
    \newcommand{\disc}{\ensuremath{_{\mathrm{disc}}}}
    \newcommand{\sink}{\ensuremath{_{\mathrm{sink}}}}
    \newcommand{\roche}{\ensuremath{_{\mathrm{Roche}}}}
    
    \newcommand{\MBH}{\ensuremath{M_{\mathrm{BH}}}}
    \newcommand{\core}{\ensuremath{_{\mathrm{c}}}}
    \newcommand{\gas}{\ensuremath{_{\mathrm{g}}}}
    \newcommand{\rhocut}{\ensuremath{\rho_{\mathrm{cut}}}}
    \newcommand{\Esp}{\ensuremath{E_{\mathrm{S}}}}
    \newcommand{\Eint}{\ensuremath{E_{\mathrm{U}}}}
    \newcommand{\Ekin}{\ensuremath{E_{\mathrm{K}}}}
    \newcommand{\Epot}{\ensuremath{E_{\mathrm{P}}}}
    \newcommand{\Lone}{\ensuremath{{\mathrm{L_1}}}}
    \newcommand{\Ltwo}{\ensuremath{{\mathrm{L_2}}}}
    \newcommand{\unbound}{\ensuremath{_{\mathrm{unbound}}}}
    \newcommand{\thydr}{\ensuremath{\tau_{\mathrm{hydr}}}}
    \newcommand{\BH}{\ensuremath{_{\mathrm{BH}}}}
    \newcommand{\AngV}{\ensuremath{\mathit{\Omega}}}
    \newcommand{\rasio}{\ensuremath{\mathrm{Rasio}}}
    \newcommand{\hlx}{\mbox{HLX-1}}
    \newcommand{\unsim}{\mathord{\sim}}
    \newcommand{\tot}{\ensuremath{_\mathrm{tot}}}

    \newcommand{\mesa}{\textsc{mesa}}
    \newcommand{\amuse}{\textsc{amuse}}
    \newcommand{\seba}{\textsc{seba}}
    \newcommand{\ficode}{\textsc{fi}}

    \newcommand\T{\rule{0pt}{2.6ex}}
    \newcommand\B{\rule[-1.2ex]{0pt}{0pt}}

\begin{document}
    \title[Eccentric mass transfer to an IMBH.]
        {Simulations of the tidal interaction and mass transfer of a star in an eccentric orbit around an intermediate-mass black hole: the case of HLX-1}

    \author[E. van der Helm et al.]
        {Edwin van der Helm$^1$,
        Simon Portegies Zwart$^1$
        and Onno Pols$^2$ \\
        $^1$Leiden Observatory, The Netherlands \\
        $^2$Radboud Universiteit Nijmegen, Dept. of Astrophysics/IMAPP, The Netherlands
        }

    \maketitle

    \begin{abstract}
        The X-ray source \hlx\ near the spiral galaxy ESO 243-49 is currently the best intermediate-mass black hole candidate.
        It has a peak bolometric luminosity of $10^{42}$\,erg\,s$^{-1}$, which implies a mass inflow rate of $\unsim10^{-4}$\,\MSun\,yr$^{-1}$, but the origin of this mass is unknown.
        It has been proposed that there is a star on an eccentric orbit around the black hole which transfers mass at pericentre.
        To investigate the orbital evolution of this system, we perform stellar evolution simulations using \mesa\ and SPH simulations of a stellar orbit around an intermediate-mass black hole using \ficode.
        We run and couple these simulations using the \amuse\ framework.
        We find that mass is lost through both the first and second Lagrange points and that there is a delay of up to 10\,days between the pericentre passage and the peak mass loss event.
        The orbital evolution timescales we find in our simulations are larger than what is predicted by analytical models, but these models fall within the errors of our results.
        Despite the fast orbital evolution, we are unable to reproduce the observed change in outburst period.
        We conclude that the change in the stellar orbit with the system parameters investigated here is unable to account for all observed features of HLX-1.
    \end{abstract}

    \begin{keywords}
        X-rays: individual (\hlx) -- X-rays: binaries -- binaries: close -- stars: kinematics and dynamics -- methods: numerical -- hydrodynamics
    \end{keywords}

\section{Introduction}
    The X-ray source \hlx\ has an estimated peak bolometric isotropic luminosity of $\unsim 10^{42}$\,erg\,s$^{-1}$ \citep{farrell_intermediate-mass_2009}.
    With redshift measurements of the optical counterpart, \citet{wiersema_redshift_2010} and \citet{soria_kinematics_2013} argue that \hlx\ coincides with the spiral galaxy ESO 243-49, but it does not coincide with its nucleus.
    The source transits in a few days from the low/hard X-ray state to the high/soft X-ray state during which the count rate increases by an order of magnitude \citep{godet_first_2009,servillat_x-ray_2011}.
    During this transition \citet{webb_radio_2012} have detected radio flares.
    The X-ray spectrum has been fitted using black hole accretion disc spectral models \citep{davis_cool_2011,godet_investigating_2012,straub_investigating_2014}.
    These arguments have been used to claim that \hlx\ hosts a black hole with a mass between $\unsim 10^4$ and $\unsim 10^5$ \MSun\ \citep{webb_radio_2012}.
    This mass range is consistent with an intermediate-mass black hole (IMBH) \citep[e.g.][]{miller_intermediate-mass_2004}.

    The luminosity of \hlx\ follows a fast rise and exponential decay pattern with a period of approximately $370$ days \citep{lasota_origin_2011}.
    Various scenarios have been investigated to explain this observed X-ray light curve.
    They postulate a star on an eccentric orbit around an IMBH.
    This star would experience Roche-lobe overflow at pericentre, transferring mass to the black hole.
    The transferred mass would then form an accretion disc and the mass accreting onto the black hole would emit the X-rays \citep{shakura_black_1973}.
    The observed $370$ day periodicity in the X-rays can then be interpreted as the orbital period of this companion star around the black hole.

    To date, 5 complete outbursts of \hlx\ have been observed with the \textit{Swift} X-ray telescope \citep{godet_investigating_2012,godet_implications_2014}.
    An overview of the time evolution of the shape of the outbursts is presented by \citet{miller_wind_2014}.
    The peak X-ray luminosity decreases with each outburst and the decay time also decreases, which means that the integrated energy per outburst decreases.
    In the proposed mass transfer models, these observations would correspond to a stable semimajor axes ($a$) and a decrease of the mass transfer rate ($\dot{M}$).

    In 2013, the expected outburst occurred more than a month later than predicted from the previously observed outburst period \citep{godet_implications_2014}.
    The outburst after that occurred in 2015 with an even larger delay \citep{kong_new_2015}
    To understand the first delay, \citet{godet_implications_2014} performed SPH simulations of the tidal capture of a star \citep[e.g.][]{baumgardt_tidal_2006} onto a highly eccentric ($e>0.9998$) orbit around a black hole.
    In their scenario, the pericentre passage induces oscillations in the star and the phase of these oscillations at the next pericentre passage affect the tidal forces.
    As this phase is essentially random the tidal forces can induce stochastic variations in the orbital period \citep{ivanov_tidal_2004}.
    These stochastic variations can explain the observed delay under the assumption that viscous processes do not damp the oscillations between pericentre passages.
    In this tidal capture model, the star will only orbit the black hole for $\unsim 10^2$\,years before it is expelled again.

    An alternative scenario to explain the observed X-ray light curve has been proposed by \citet{miller_wind_2014}.
    They postulate an encounter between the IMBH and a high mass giant star between 5 and 15\,years ago.
    In this encounter, the envelope of the giant star would have been stripped by tidal interaction with the black hole.
    The core of the giant would have remained on an eccentric orbit around the black hole.
    This core would have a low mass hydrogen envelope which it loses through a strong stellar wind.
    At every pericentre passage this wind feeds the accretion disc and at that point this scenario follows the same arguments as the scenario proposed by \citet{lasota_origin_2011}.
    This second scenario predicts that the stripped envelope material will be accreted onto the black hole resulting in a stable bright X-ray signal within $10-100$\,years.
    From the current observations, \citet{miller_wind_2014} have not been able to exclude either scenario for \hlx.

    The observations of \hlx\ indicate that the black hole has an accretion disc which has been studied by fitting disc models to the observed spectra.
    The disc is cool \citep{davis_cool_2011} and thin \citep{godet_investigating_2012} and has an outer radius between 15 and 150\,\RSun \citep{soria_eccentricity_2013}.
    It has not been possible to directly constrain the nature or the orbit of the stellar companion of \hlx\ from the observations, but the periodicity of the X-ray signal is interpreted as periodic mass transfer at pericentre, which would imply that the orbit is not circular.
    By assuming that the outer radius of the accretion disc corresponds to the pericentre distance of the stellar orbit, \citet{soria_eccentricity_2013} derive an eccentricity of $e \approx 0.95$, with a lower limit of $e \approx 0.9$.

    The direct mass transfer model has been proposed \citep{lasota_origin_2011} to explain the current observations.
    If we assume that the direct mass transfer model is correct, then the question remains: What is the origin and long term orbital evolution of the stellar companion?
    In this paper we combine SPH simulations and stellar evolution models to study the orbital evolution of the star during mass transfer.
    We then compare the results of these simulations with analytical predictions of this orbital evolution.
    Using the orbital evolution from our simulations, we will attempt to constrain the origin and long term orbital evolution of the stellar companion of \hlx.

\section{Setting the scene} \label{sec: physical parameters}
        \begin{figure*}
            \begin{center}
            \includegraphics[width=\textwidth]{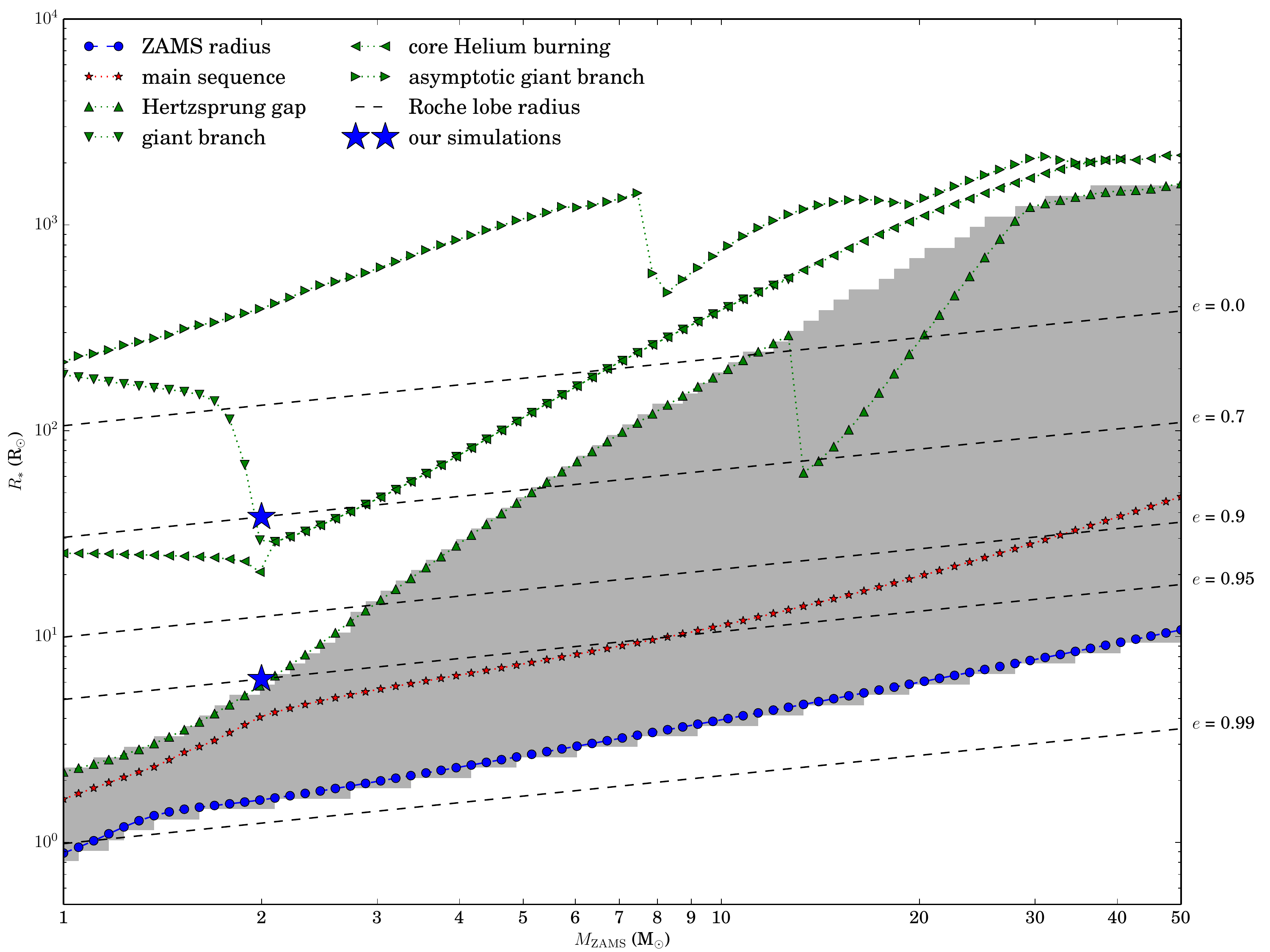}
            \end{center}
            \caption{
                The maximum radii of stars during various stages in the stellar evolution as a function of initial mass using \seba\ \citep{portegies_zwart_seba:_2012} with metallicity $Z = 0.02$.
                The zero-age main-sequence (ZAMS) radii represent the start of the main-sequence.
                The black dashed lines correspond to the Roche-lobe radius of the stellar companion of \hlx\ for each mass.
                Each line corresponds to a given eccentricity, shown on the right hand side of the figure.
                Stars in the grey shaded region have a radiation dominated stellar envelope while all other stars have a convection dominated envelope.
                The large blue stars represent the parameters used in our simulations.
                For a 2 \MSun\ star, mass transfer occurs during the first asymptotic giant branch if $e = 0.7$, or at the start of the first giant branch if $e = 0.95$.
                \label{fig: stellar radius}}
        \end{figure*}

        There is a general agreement between independent measurements of the black hole mass (\MBH) using Eddington scaling \citep[$\MBH > 9\cdot10^3 \; \MSun$,][]{servillat_x-ray_2011}, accretion disc models \citep[$\MBH \approx 10^{3-5} \; \MSun$,][]{davis_cool_2011,godet_investigating_2012,straub_investigating_2014} and the detection of ballistic jets \citep[$\MBH \approx 10^{4-5} \; \MSun$,][]{webb_radio_2012}.
        Here we will adopt $\MBH = 10 \, 000\,\MSun$ based on the assumption that at the peak of the outbursts, the IMBH luminosity is near the Eddington luminosity.
        Following the direct mass transfer model, we assume that the orbital period is equal to the observed periodicity of the outbursts ($370$\,days).
        Using the equations for a Keplerian orbit, the semimajor axis can be calculated from the orbital period and total mass of the system ($a = 21.7$\,AU) where the total mass is essentially \MBH.
        The initial parameters used in this work are summarized in tables~\ref{tab: fixed simulation parameters}~and~\ref{tab: varied simulation parameters}.

        For the outer radius of the accretion disc ($r\disc$) we use the largest value from observations, so $r\disc = 150$\,\RSun\ \citep{soria_eccentricity_2013}.
        If the pericentre distance (\rp) corresponds to the outer radius of the accretion disc, then $e \approx 0.95$ \citep{soria_eccentricity_2013}.
        The outer radius of the accretion disc does not necessarily correspond to \rp, other factors in the complex physics of accretion discs can limit the outer radius.
        We therefore investigate a lower value of $e = 0.7$, as well as $e=0.95$.
        The stellar angular velocity ($\AngV_*$) is not constrained by the observations, we therefore perform simulations with 0, 0.5 and 1 times the orbital angular velocity at pericentre ($\Omega\orb$).

    \subsection{The donor star}
        To determine the stellar radius ($R_*$) we start with the volume equivalent Roche radius for a star in an eccentric binary ($R\roche$).
        If $R_* < R\roche$ no mass transfer occurs but tidal interaction will affect the shape of the donor as well as the orbit.
        If $R_* \gtrsim R\roche$ mass will flow from the donor to the black hole, which is the case of interest in this study.
        We therefore adopt $R_* \approx R\roche$ in our calculations.
        Determining the value of $R\roche$ as a function of the mass ratio and orbital parameters requires numerical evaluation of the gravitational potential.
        In this paper, we use fitting formulas for $R\roche$ \citep{eggleton_approximations_1983,sepinsky_equipotential_2007}.
        We perform a series of simulations using $0.8 < R_*/R\roche < 1.2$ to investigate the effect of different mass loss rates on the orbital evolution.

        The stage of stellar evolution that the star is in when $R_* \approx R\roche$ depends on the zero-age main-sequence mass ($M\zams$) of the star and the orbital parameters.
        In figure~\ref{fig: stellar radius} we compare $R_*$ at different stellar evolution stages with $R\roche$ for various values of $e$ and $M\zams$ with metallicity $Z=0.02$.
        The star can be on the main-sequence when $R_* = R\roche$ if it has a high eccentricity and large mass ($e \gtrsim 0.95$ and $M\zams \gtrsim 8\,\MSun$).
        The large blue stars in figure \ref{fig: stellar radius} correspond to the orbital and stellar parameters used in this work.
        The star we use in our simulations is a giant with a convective envelope.

        The stellar mass determines the stellar lifetime; therefore the possible mass range of the stellar companion depends on the age of the stellar population near \hlx.
        Observations favour a young ($\unsim 20$\,Myr) stellar population, but an additional older population cannot be excluded \citep{farrell_combined_2014}.
        Because the companion star can originate from either population, there is only a limited observational constraint on the mass $M_*$ of the companion star.
        We adopt $M\zams = 2\,\MSun$.

    \subsection{Orbital evolution processes} \label{sec: orb evolution processes}
        Before and during mass transfer, the star orbiting the IMBH is subject to tidal dissipation.
        The effects of tidal dissipation on the orbit of a star with a convective envelope are commonly described by the equilibrium tide model \citep{zahn_tidal_1977,hut_tidal_1981,rasio_tidal_1996,eggleton_equilibrium_1998}.
        As a result of the tidal dissipation, the orbit gradually becomes circular ($e = 0$) and the star will evolve towards corotation.
        Corotation is defined as the state where the stellar angular velocity equals the orbital angular velocity ($\AngV_*= \Omega_{\mathrm{orb}}$).
        The time-scale for establishing corotation is shorter than the time-scale for circularization, it is therefore possible to define a state of pseudo-synchronization, where $\dot{\AngV_*} = 0$ for the current value of $e$.
        When the system is in pseudo-synchronization, $\AngV_*$ is close to the orbital angular velocity at pericentre ($\AngV_* > 0.799 \, \Omega\orb$) \citep{hut_tidal_1981}.
        The rate of change of the orbital and stellar parameters depends on the stellar radius as $\dot{a} \propto R_*^8$, $\dot{e} \propto R_*^8$ and $\dot{\AngV_*} \propto R_*^6$.
        The tidal dissipation is stronger when the star has a convective envelope than when it has a radiative envelope, in which case the equilibrium tide model is inadequate.

        The main uncertainty in the equilibrium tide model is the tidal damping time-scale ($T$) which is usually estimated in combination with the apsidal motion constant ($k$) because the rates of change in the tidal evolution equations scale with $k/T$.
        The constant $k$ is a measure of the central condensation of the star, where lower $k$ corresponds to a higher central condensation \citep[e.g.][]{hut_tidal_1981}.
        A theoretical estimate of $k/T$ as a function of the stellar parameters has been calculated by \citet{rasio_tidal_1996}.
        However, comparisons with observations indicate that $k/T$ should be larger \citep[e.g.][]{meibom_robust_2005,belczynski_compact_2008}.
        Using the equations of \citet{hut_tidal_1981} and \citet{rasio_tidal_1996} and the observed parameters of \hlx\ (with $R_* = R\roche$), we find that, due to tidal dissipation, $\AngV_*$ changes on a time-scale of $\unsim10^{2-4}$\,years, $a$ changes on a time-scale of $\unsim10^{6}$\,years and $e$ changes on a time-scale of $\unsim10^{7-8}$\,years.

        If a star in a binary loses mass the orbit changes due to the change in the mass and angular momentum of the star.
        The lost mass can be transferred from the star to the other object, and part of the mass can be ejected from the system.
        The evolution of the orbit due to mass transfer when $e>0$ has been studied analytically \citep{sepinsky_interacting_2007,sepinsky_interacting_2009}.
        They calculate the change in the semimajor axis $a$ and the eccentricity $e$ as a function of $\dot{M}_*$.
        They assume that the mass is lost or transferred instantaneously during the pericentre passage, and that all mass leaves the star through the first Lagrangian point \Lone\ of the eccentric orbit.
        Using these equations and the observed parameters of \hlx, we find that $a$ changes on a time-scale of $\unsim10^{4-5}$\,years and $e$ changes on a time-scale of $\unsim10^{5-6}$\,years due to mass transfer.
        Mass transfer in eccentric binary systems has also been studied using smoothed particle hydrodynamics (SPH) \citep{regos_mass_2005,church_mass_2009,lajoie_mass_2011} but they did not investigate the effect on the orbit.

        The effect of mass transfer from a star in an eccentric orbit around a super massive black hole has also been studied using grid based hydrodynamics simulations with adaptive mesh refinement \citep{macleod_spoon-feeding_2013}.
        They assumed in their models that the stellar orbit does not change through mass transfer or tidal interactions.
        They argue that this assumption is valid when the stellar specific orbital energy ($E_{*,\mathrm{orb}} \propto \MBH/a$) is larger than the stellar specific binding energy ($E_{*,\mathrm{bind}} \propto M_*/R_*$).
        This would imply that $a$ is constant, which is not the case for our system (see table~\ref{tab: timescales}).
        Furthermore, even if $a$ where constant, changes in $e$ could still affect the mass transfer.

        Relativistic effects can affect the orbit of a star near a black hole.
        The first relativistic correction that causes a change in $a$ and $e$ is the emission of gravitational waves.
        The magnitude of this effect can be calculated using the equation from \citet{peters_gravitational_1964}, which results in an orbital evolution time-scale of $\unsim10^{10-11}$\,years.
        This is considerably longer than the orbital evolution caused by other effects and therefore we conclude that relativistic effects can be ignored in this work.

        \begin{table}[t]
            \caption{
                An overview of the time-scales ($\tau$) of the orbital and stellar processes in \hlx.
                For a parameter $X$, the time-scale $\tau_X$ is defined here as $|X/\dot{X}|$.
                The second column contains the affected parameter and the third column contains the literature source for the value or analytical formula.
                Where applicable, system parameters from section~\ref{sec: physical parameters} have been used.
                \label{tab: timescales}}

            \begin{tabular}{l r l l }
                \B process & & $\tau$ (yr) & source \\
                \hline
                \T tidal effects & $\AngV_*$ & $10^{2-4}$ & \citet{hut_tidal_1981} \\
                & $a$ & $10^{6}$ & \citet{rasio_tidal_1996} \\
                & $e$ & $10^{7-8}$ & \\
                mass transfer & $a$ & $10^{4-5}$ & \citet{sepinsky_interacting_2009} \\
                & $e$ & $10^{5-6}$ & \\
                Relativistic effects & $a$ & $10^{10}$ & \citet{peters_gravitational_1964} \\
                & $e$ & $10^{11}$ & \\
            \end{tabular}
        \end{table}

        In table~\ref{tab: timescales} we summarize the time-scales of the evolutionary processes discussed in this section.
        The change in $\AngV_*$ through tidal effects is faster than the change in the other orbital and stellar parameters.
        The evolution of $a$ and $e$ appears to be dominated by mass transfer rather than tidal effects but this may not be entirely correct due to the uncertainty in $k/T$.
        We therefore take both tidal effects and mass transfer into account in our simulations.

\section{Methods} \label{sec: Methods}
        We simulate the evolution of \hlx\ in two steps.
        The first step spans $\unsim 10^9$\,years, in which we simulate the stellar evolution up to the present day.
        The second step spans $\unsim 10$\,years, in which we simulate the hydrodynamical evolution of the system in detail.

        In the first step, we evolve a single star in isolation until it has a radius comparable to the Roche radius in the chosen initial orbit.
        In principle, we cannot assume that the star has evolved in isolation because the formation history of \hlx\ is unknown.
        However, if the star evolved on an orbit similar to the current orbit then $R_* < R\roche$ and the influence of the black hole is negligible during this part of the evolution.
        If the star was captured into the present orbit, it would have evolved far from the black hole, and we can also assume it evolved in isolation.

        We convert the one-dimensional stellar structure model to a three-dimensional gas model.
        We place this star in a Keplerian orbit around the black hole using \MBH\ and orbital parameters from section~\ref{sec: physical parameters}.
        This complete system is then used as the initial condition for the second and most important part of our simulations.

        In the second step, we simulate the detailed three-dimensional gas and gravitational dynamics of the system.
        The mass transfer resulting from the gravitational interaction is measured and we calculate the orbital mechanics of all gas.
        We then calculate the secular change in the orbit of all gas that is bound to the star.
        The time for which we simulate this step is comparable to the time for which \hlx\ has been observed.
        We have also performed a number of longer simulations to investigate the change in $\AngV_*$ through tidal interaction.

    \subsection{Numeric implementation} \label{sec: numeric implementation}
        We use the \amuse\footnote{amusecode.org} framework \citep{portegies_zwart_multi-physics_2013,pelupessy_astrophysical_2013,van_elteren_multi-scale_2014} to perform all simulations in this work.
        For the evolution of a single isolated star, we use \mesa\ \citep{paxton_modules_2011} as it is implemented in \amuse.
        \mesa\ is a one-dimensional Henyey code which can be used to perform stellar evolution calculations by assuming spherical symmetry and hydrostatic equilibrium.
        For the three-dimensional gas-dynamical simulations we use \ficode\ \citep{pelupessy_numerical_2005} as it is implemented in \amuse.
        \ficode\ is an SPH \citep{monaghan_smoothed_1992} code in which the gas is represented by discrete particles.

        To create the three-dimensional gas model of the star, we take the radial density, temperature and mean molecular mass profiles from the \mesa\ model and generate a set of SPH particles.
        This is done using the \amuse\ routine \texttt{star\_to\_sph.py} \citep{de_vries_evolution_2013}.
        Accurately simulating high density regions requires small time-steps in SPH simulations, simulating the core of the star therefore requires more CPU time than the envelope.
        However, to study mass transfer and tidal interactions we mainly have to resolve the outer parts of the star and not the core.
        We therefore replace the stellar core with a single mass point and prevent the star from collapsing by adding Plummer softening to the core particle .
        The softening length depends on the mass of the core particle and the original entropy profile \citep{de_vries_evolution_2013}.

        The IMBH is represented by a single point mass in the SPH code, which is also a sink particle \citep{bate_modelling_1995}.
        At every step in the simulation, any SPH particle within a radius $r\sink$ from the black hole is accreted, which is implemented in the \texttt{sink.py} routine in \amuse.
        We do not resolve the accretion disc but we set the radius of the black hole sink particle equal to the radius of the accretion disc ($r\sink = r\disc$).

    \subsection{Relaxation} \label{sec: relaxation}
        Part of the SPH method is that the kernel function induces a non-physical force that prevents the SPH particles from approaching each other arbitrarily closely \citep[e.g.][]{price_smoothed_2012}.
        Particles can start close together because the initial spatial distribution of SPH particles is chosen randomly and therefore this non-physical force can add energy and cause the star to expand, which is not physical.
        To prevent this from happening, the gas has to be relaxed in the correct gravitational potential before we start the simulation.
        Because the gravitational potential is not static in an eccentric orbit, we perform this relaxation in multiple steps.

        In the first step, the particle distribution is brought to dynamical equilibrium in the fixed gravitational potential at apocentre in orbit around the black hole.
        For this step we use the relaxation method described in section 3.3 of \citet{de_vries_evolution_2013}.
        We evolve the SPH system by a single timestep and subtract the bulk motion of the star to preserve the centre-of-mass position and velocity.
        Then we decrease the change in velocity by multiplying it with a factor $f$ that increases linearly from 0 to 1 over 400\,days (approximately one orbit).

        In the second step, the star is evolved in a circular orbit with a semimajor axis (relaxation distance) $a_{rel} = 10$ AU.
        We choose this value because the star is close to the black hole but not close enough for mass transfer to occur.
        After one orbit, the relaxed star is returned to the apocentre position of the original orbit.

    \subsection{Orbital parameter determination} \label{sec: orbital parmaters}
        To calculate the (change in) orbital parameters, we need the mass, position and velocity of the black hole and the star.
        For the black hole, we know the position and velocity because it is a single particle in our simulation.
        For the star, we determine which SPH particles are bound to the stellar core particle and to each other, and then compute the total mass, the centre of mass position and velocity of these particles.

        To determine the bound particles, we calculate the specific total energy (\Esp) from the internal (\Eint), kinetic (\Ekin) and potential (\Epot) energy of each particle.
        The internal energy is calculated within the SPH code.
        The kinetic and potential energy of each particle are calculated from the mass ($M\core$), position ($\vec{X}\core$) and velocity ($\vec{V}\core$) of the stellar core, and the position ($\vec{X}\gas$) and velocity ($\vec{V}\gas$) of the gas particle.
        \begin{align}
            \Esp & = \Eint + \Ekin - \Epot \\
            \Ekin & = \frac{1}{2} \left ( \vec{V}\gas - \vec{V}\core \right )^2 \\
            \Epot & = \frac{G * M\core}{ \left | \vec{X}\gas - \vec{X}\core \right |}
        \end{align}
        Here $G$ is the gravitational constant.
        We consider a gas particle bound to the stellar core when $\Esp < 0$.

        This set does not include all bound particles yet, as we have not taken the gravitational binding energy between gas particles into account.
        We therefore replace $M\core$, $\vec{X}\core$ and $\vec{V}\core$ with the total mass, the centre of mass and the centre of mass velocity of the stellar core plus the known bound particles and recalculate \Esp.
        This procedure is repeated until no more particles are added in the next iteration.

        In the calculations for a Keplerian orbit it is assumed that the black hole and the star are point particles.
        This assumption is valid while the radius of the star is far smaller than the orbital separation.
        Because this assumption is not valid near pericentre, the computed orbital parameters are not constant throughout the orbit.
        We calculate all orbital evolution trends from the orbital parameters as measured at apocentre.

    \subsection{Stellar radius determination} \label{sec: stellar radius}
        \begin{figure}
            \includegraphics[width=0.5\textwidth]{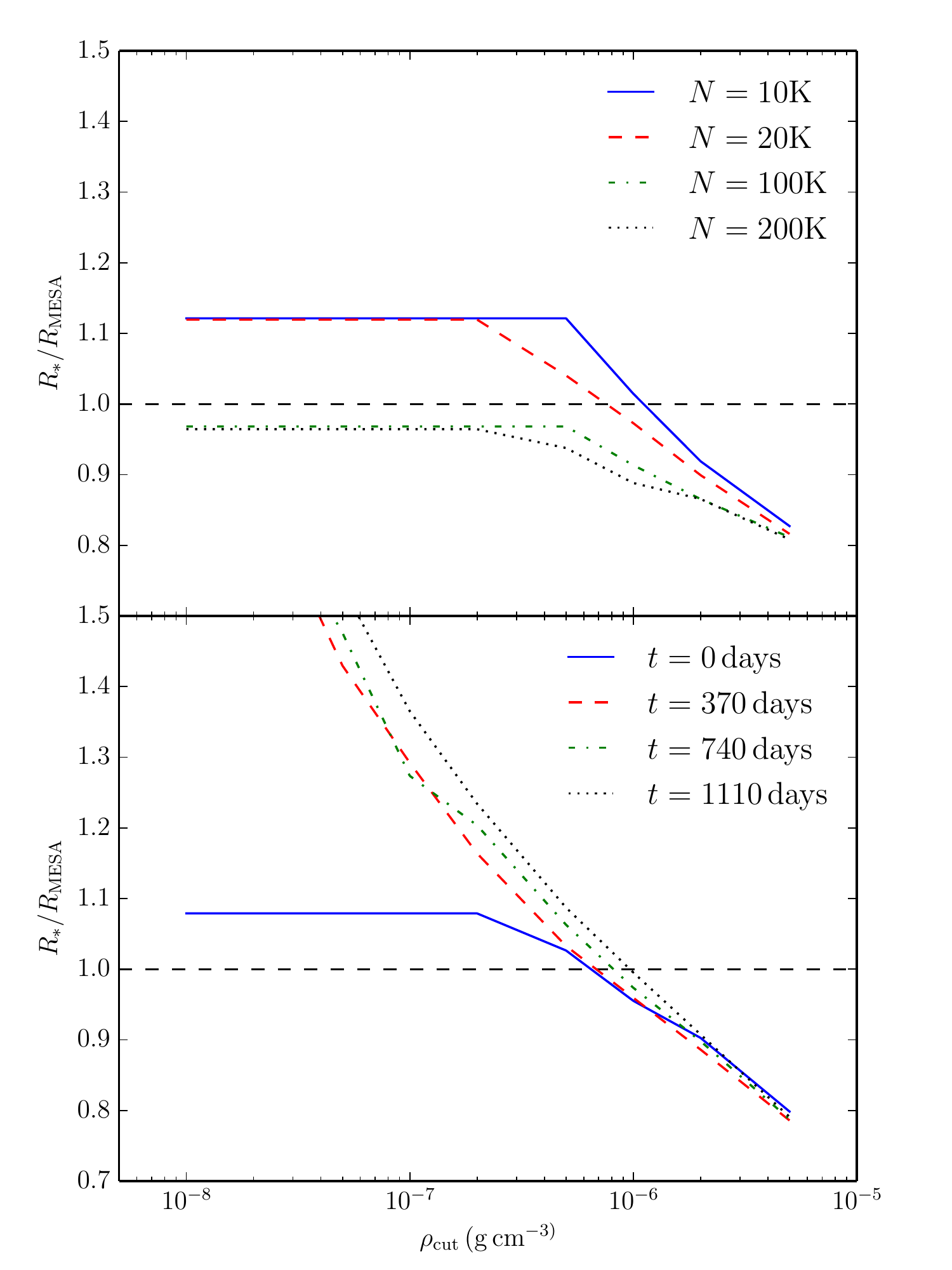}
            \caption{
                The radius of the SPH realization of the star calculated with different values of the density cut-off (\rhocut) for different resolutions (top panel) and different times in the simulation (bottom panel).
                The radius of the \mesa\ model is shown with dashed black line, we find good agreement with the \mesa\ model for $\rhocut = 5\times10^{-7}$\,g\,cm$^{-3}$.
                If \rhocut\ is smaller than this value, the measured radius at later times in the simulation is very large and completely dominated by a few SPH particles that are only barely bound to the star.
                \label{fig: radius density cutoff}}
        \end{figure}

        Both the rate of tidal evolution of the binary system and the rate of mass transfer are very sensitive to the radius of the donor star.
        In order to make a comparison between the simulations and analytical prescriptions we need to have a good estimate of the stellar radius from the simulations.
        In SPH simulations it is hard to acquire such an estimate because the star is represented by a set of discrete particles.
        We have experimented with various methods to measure the radius, such as using the outermost  bound particle or using the mean distance from the centre of mass of the $n$ outermost SPH particles.
        However, we found that these methods are far too sensitive to a small number of barely bound SPH particles in the outer regions of the star.
        We solve this problem by introducing a density cut-off (\rhocut) and we measure the stellar radius at this density.

        In figure~\ref{fig: radius density cutoff} we present the measured stellar radius for different values of \rhocut.
        The radius at a fixed density is smaller for higher resolution and larger at a later time in the simulation.
        When we adopt a density cut-off of $5\times10^{-7}$\,g\,cm$^{-3}$ the SPH radius determination is consistent (within 5\,per\,cent) of the results from the stellar evolution code for the resolution used in this work.

    \subsection{Model parameters} \label{sec: model parameters}
        \begin{figure}
            \includegraphics[width=0.5\textwidth]{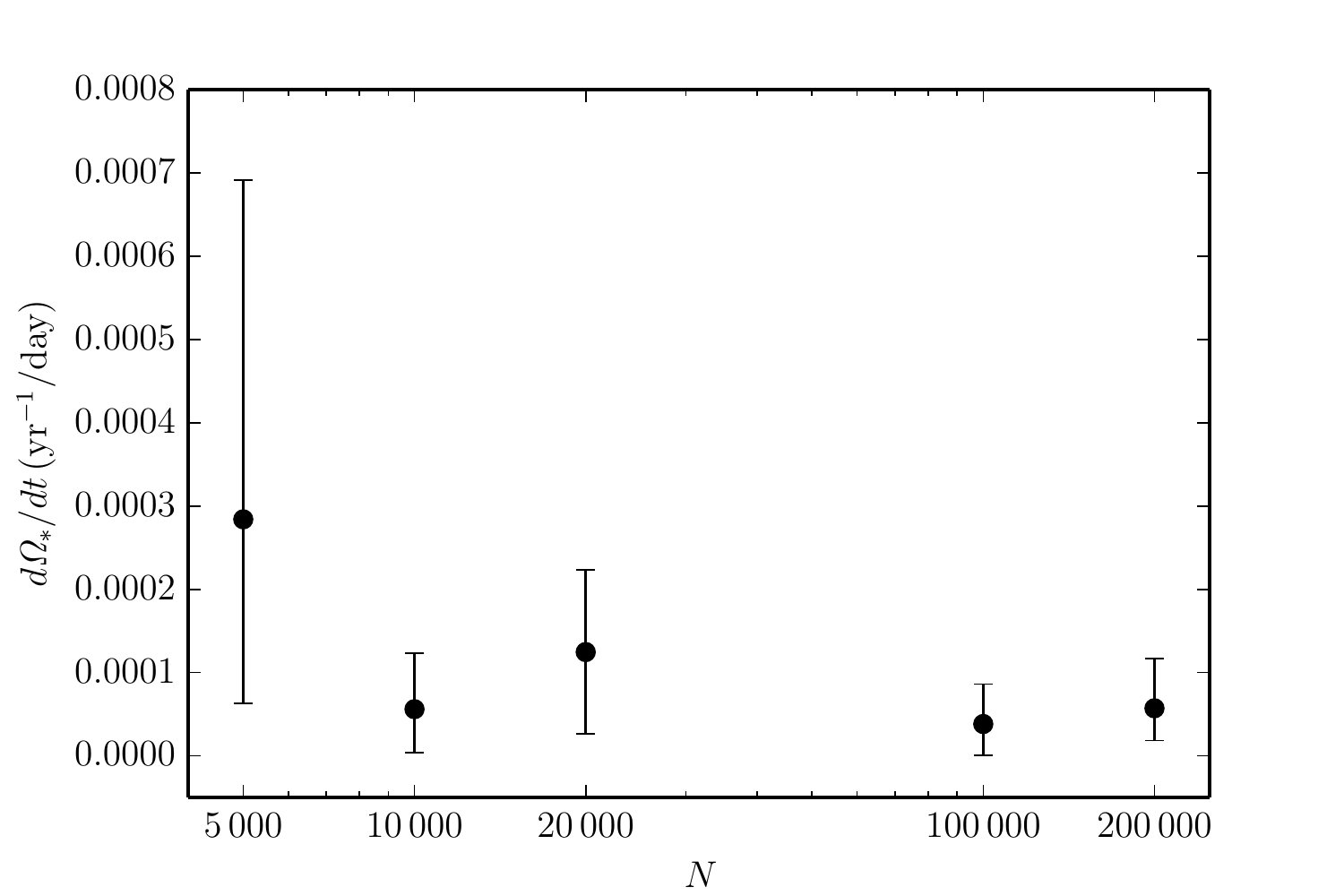}
            \caption{
                The change in the stellar angular velocity while the star is far away from the black hole (the distance is greater than the semimajor-axis) as a function of resolution.
                $d\AngV_*/dt$ should be 0 this far from the black hole and we see that for $N > 10\,000$ a larger $N$ does not result in a smaller $d\AngV/dt$.
                \label{fig: error convergence}}
        \end{figure}

        \begin{figure}
            \includegraphics[width=0.5\textwidth]{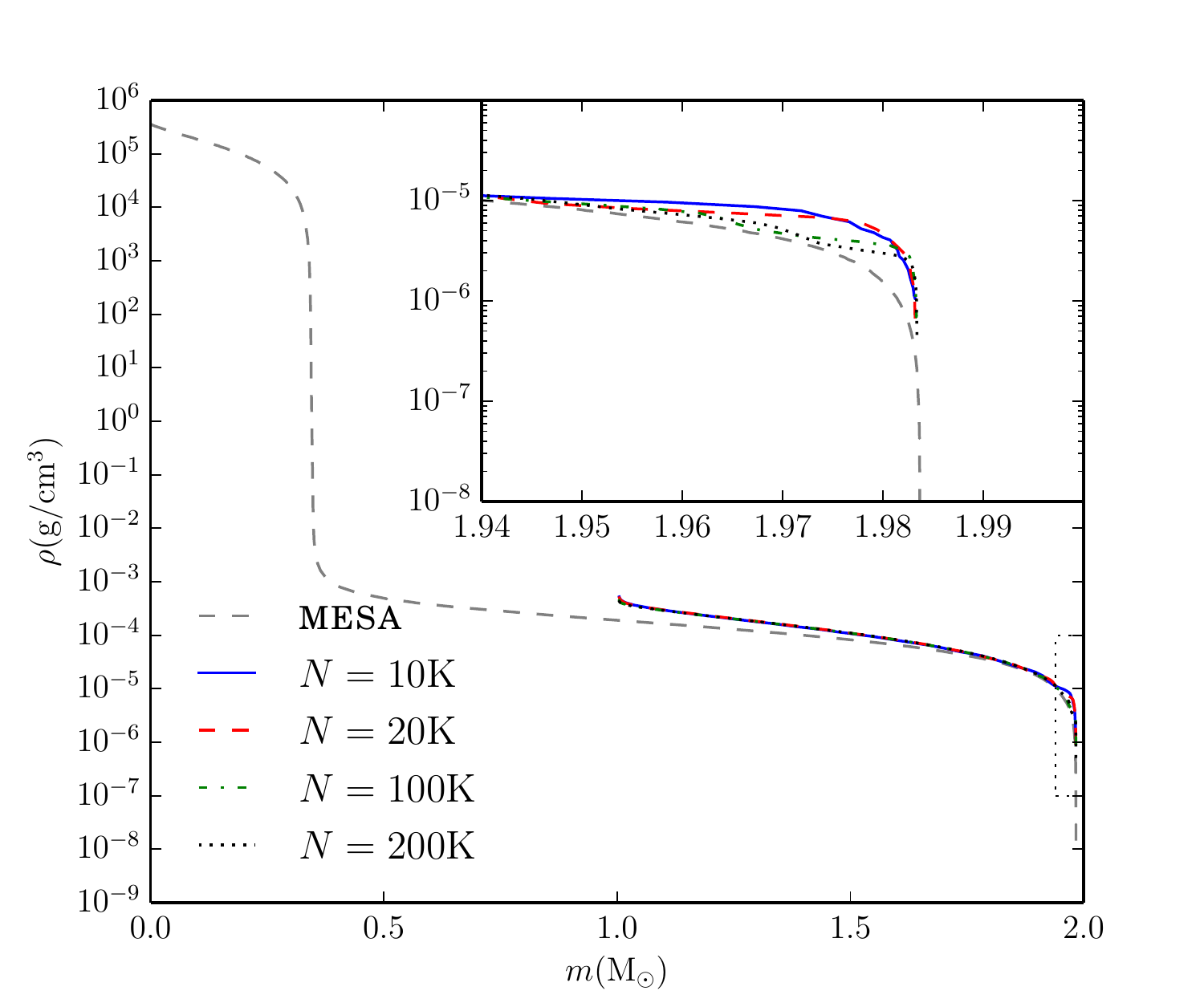}
            \caption{
                The radial density profile of the \mesa\ model compared to the density of the SPH particles that represent it for different values of $N$.
                The core 1\,\MSun\ of the star is not included because it is not represented by SPH particles.
                It can be seen that models with more SPH particles have a lower density in the outermost layers of the star.
                \label{fig: radial density}}
        \end{figure}

        Using a computer simulation to approximate a physical process introduces a number of assumptions that are generally characterized by free parameters.
        Here we summarize the assumptions used in this work and test the effect of varying these free parameters on our results.

        For the stellar evolution simulations we assume a metallicity $Z = 0.02$, but the actual metallicity of the star is unknown.
        The stellar wind mass-loss follows the Reimers mass-loss model with an efficiency $\eta_R = 0.5$ \citep{reimers_circumstellar_1975}.
        We do not include convective overshooting in the stellar evolution simulations.

        For the SPH simulations we assume an adiabatic equation of state without cooling.
        While cooling can affect the mass transfer rate, we suspect that the influence of cooling on our results will be small and we do not investigate it.
        We use an adaptive smoothing length with the number of neighbors set to 64 \citep{pelupessy_numerical_2005}.

        The resolution of the simulation depends on the number of SPH particles ($N$).
        The mass of individual SPH particles is inversely proportional to their number in the sense that more particles result in a higher resolution but this comes at the expense of more computer time.
        We search for a balance between cost and resolution using convergence tests.
        We base our convergence test on $\AngV_*$ because this parameter is important for the tidal interactions discussed in section~\ref{sec: orb evolution processes}.

        For the main convergence test, we require that the angular velocity of the donor star ($\AngV_*$) is stable when the star does not interact with the black hole.
        Characterizing the stellar rotation rate by a single value $\AngV_*$ is not always possible because different parts of the star can have different rotational velocities.
        However, we have found that the star in our simulations generally rotates as a rigid body when it is far away from the black hole, so we can take $\AngV_*$ to be the average of $\AngV$ of all the SPH particles that represent the star.
        For this reason, we measure the rotation rate of the star in the binary when the distance between the black hole and the star is larger than the semi-major axis.
        For simulations with the star is initially not rotating, $\AngV_*$ increases at every pericentre passage.
        In between pericentre passages, the rotation of the star should remain constant, but this turns out to depend on the resolution of the simulation.
        By performing several simulations with various resolutions we find that for $N \gtrsim 10\,000$, the erroneous change in $\AngV_*$ does not depend on $N$ (figure~\ref{fig: error convergence}).
        We conclude that the solution is converged and we use $N = 20\,000$.

        At the resolution we have chosen, we still have a slight mismatch in the outermost density profile of the SPH realization compared to the underlying stellar evolution model (see figure~\ref{fig: radial density}).
        This resolution therefore also does not result in a converged stellar radius compared to the stellar evolution model on which the initial realization was based (see figure~\ref{fig: error convergence}).
        Using a resolution that is high enough to ensure that the outermost density profile and the stellar radius are converged is not computationally feasible at this time.
        Instead we take this limitation into account when interpreting our results.
        In figure~\ref{fig: radial density} we also see that gas near the core particle has a higher density than the gas in the underlying stellar evolution model.
        Because the effects studied in this paper are dominated by the behaviour of the outer layers of the star, we do not expect this discrepancy to affect our results.

        An artificial viscosity \citep[$\alpha$ and $\beta$, see][]{monaghan_smoothed_1992} is required in SPH to model discontinuities such as shocks.
        The dimensionless parameters can be chosen between $\alpha=\beta=0$ (no artificial viscosity) and $\alpha=1;\beta=2$ (artificial viscosity proportional to the resolution length).
        For simulations with strong shocks, an adaptive artificial viscosity can be used to correctly model instabilities \citep[e.g.][]{morris_switch_1997}.
        We do not expect strong shocks in our simulations so a fixed artificial viscosity is sufficient and we adopt $\alpha = 0.5;\beta=1.0$ as these are the optimal values for most problems \citep{lombardi_tests_1999}.
        The value of $\alpha$ can influence the effective $k/T$ in our simulations, and therefore we also perform simulations with $\alpha = 0.1$ and $\alpha = 0.01$ (and $\beta = 2 \alpha$) to quantify this effect.

        As described in section~\ref{sec: numeric implementation}, we replace the core of the star with a single core particle with mass $M\core$.
        There are two practical constraints that affect the value of $M\core$.
        If $M\core$ is too large the SPH particles that represent the star cannot reproduce the stellar density profile.
        If $M\core$ is too small, the simulation will take a long time to run while the resolution at the surface of the star is not affected.
        We have created stellar models with a number of values of $M\core$ and have found that $M\core = 0.5$\,M\zams\ provides a good balance between these constraints.

\section{Results} \label{sec: Results}
        \begin{table}
            \caption{
                The initial parameters that we did not vary between the main simulations used in this work.
                The initial parameters that we did vary between simulations can be found in table~\ref{tab: varied simulation parameters}
                \label{tab: fixed simulation parameters}}

            \begin{tabular}{lll}
                \B name & parameter & value \\
                \hline
                \T black hole mass & \MBH & 10\,000\,\MSun \\
                orbital period & $P$ & 370\,days \\
                companion mass & $M\zams$ & 2\,\MSun \\[2mm]
                semimajor axis & $a$ & 21.7\,AU \\
                acc. disc radius & $r\disc$ & 150\,\RSun \\[2mm]
                number of particles & $N$ & 20\,000 \\
                core mass & $M\core$ & $\frac{1}{2}\,M\zams$  \\
                viscosity & $\alpha$ and $\beta$ & 0.5 and 1.0 \\
            \end{tabular}
        \end{table}

        \begin{table}
            \center
            \caption{The 20 main simulations used in this work and the initial parameters that were varied between them.
                    For the three simulations in bold font, we also performed 8 additional test simulations varying $N$ and $\alpha$ (below table).
                    \label{tab: varied simulation parameters}}

            \begin{tabular}{llllllll}
                name & $e$ & $\AngV_*$ & $R_*$ & $R_*/R\roche$ & age \\
                \B & & ($\Omega\orb$) & (\RSun) & & (yr) \\
                \hline
                \T \textbf{O0R37} & 0.7  & 0   & 37  & 0.84 & 1.0360e9 \\
                O0R39    & 0.7  & 0   & 39  & 0.88 & 1.0367e9 \\
                O0R41    & 0.7  & 0   & 41  & 0.93 & 1.0373e9 \\
                O0R43    & 0.7  & 0   & 43  & 0.97 & 1.0378e9 \\
                O0R45    & 0.7  & 0   & 45  & 1.02 & 1.0383e9 \\[2mm]
                O5R37    & 0.7  & 0.5 & 37  & 0.87 & 1.0360e9 \\
                O5R39    & 0.7  & 0.5 & 39  & 0.92 & 1.0367e9 \\
                O5R41    & 0.7  & 0.5 & 41  & 0.97 & 1.0373e9 \\
                O5R43    & 0.7  & 0.5 & 43  & 1.01 & 1.0378e9 \\
                O5R45    & 0.7  & 0.5 & 45  & 1.06 & 1.0383e9 \\[2mm]
                O1R37    & 0.7  & 1   & 37  & 0.97 & 1.0360e9 \\
                \textbf{O1R39}    & 0.7  & 1   & 39  & 1.02 & 1.0367e9 \\
                O1R41    & 0.7  & 1   & 41  & 1.07 & 1.0373e9 \\
                O1R43    & 0.7  & 1   & 43  & 1.13 & 1.0378e9 \\
                O1R45    & 0.7  & 1   & 45  & 1.18 & 1.0383e9 \\[2mm]
                O1R6.0   & 0.95 & 1   & 6.0 & 0.96 & 9.7004e8 \\
                O1R6.2   & 0.95 & 1   & 6.2 & 1.00 & 9.7095e8 \\
                O1R6.4   & 0.95 & 1   & 6.4 & 1.03 & 9.7207e8 \\
                O1R6.6   & 0.95 & 1   & 6.6 & 1.06 & 9.7302e8 \\
                \textbf{O1R6.8}   & 0.95 & 1   & 6.8 & 1.09 & 9.7418e8 \\
            \end{tabular}

            \addvspace{5mm}

            \begin{tabular}{lrl}
                O0R37 & $\rightarrow$  & $N = 10\,000$, $N = 100\,000$, $N = 200\,000$ \\
                      &                & $\alpha=0.1$, $\alpha=0.01$ \\
                O1R39 & $\rightarrow$  & $N = 100\,000$, $N = 500\,000$ \\
                O1R6.8 & $\rightarrow$ & $N = 100\,000$
            \end{tabular}
        \end{table}

        We have performed 20 simulations with different initial conditions to investigate the orbital evolution of \hlx.
        In table~\ref{tab: fixed simulation parameters} we present the initial conditions that we did not vary in the 20 main simulations.
        In table~\ref{tab: varied simulation parameters} we present the initial conditions that we did vary between the 20 main simulations.
        We used 8 test simulations to investigate the effect of different values of $N$ and $\alpha$.
        The 20 main simulations are divided into 4 sets with different orbital parameters $e$ and $\AngV_*$.
        For each of these 4 sets, we have performed 5 simulations with different values of $R_*$.
        We have performed the O1R39 simulation with $N = 500\,000$ twice to investigate an unexpected effect of the numerical error in the angular momentum (see section \ref{sec: Orbital evolution}).
        The age at which the star reaches the desired radius is also included in table~\ref{tab: varied simulation parameters}.

    \subsection{Tidal dissipation} \label{sec: Tidal dissipation}
        \begin{figure*}
            \includegraphics[width=0.49\textwidth]{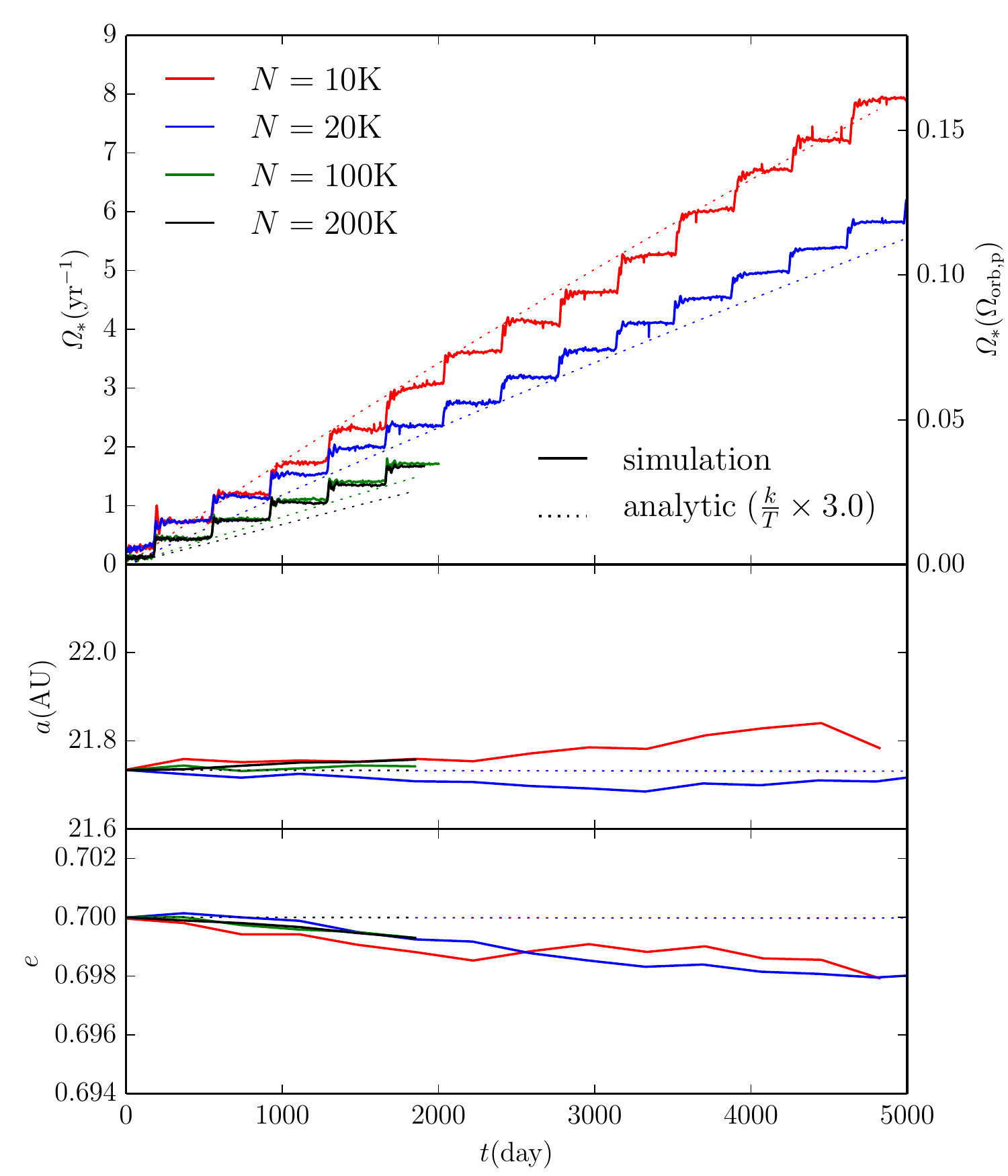}
            \includegraphics[width=0.49\textwidth]{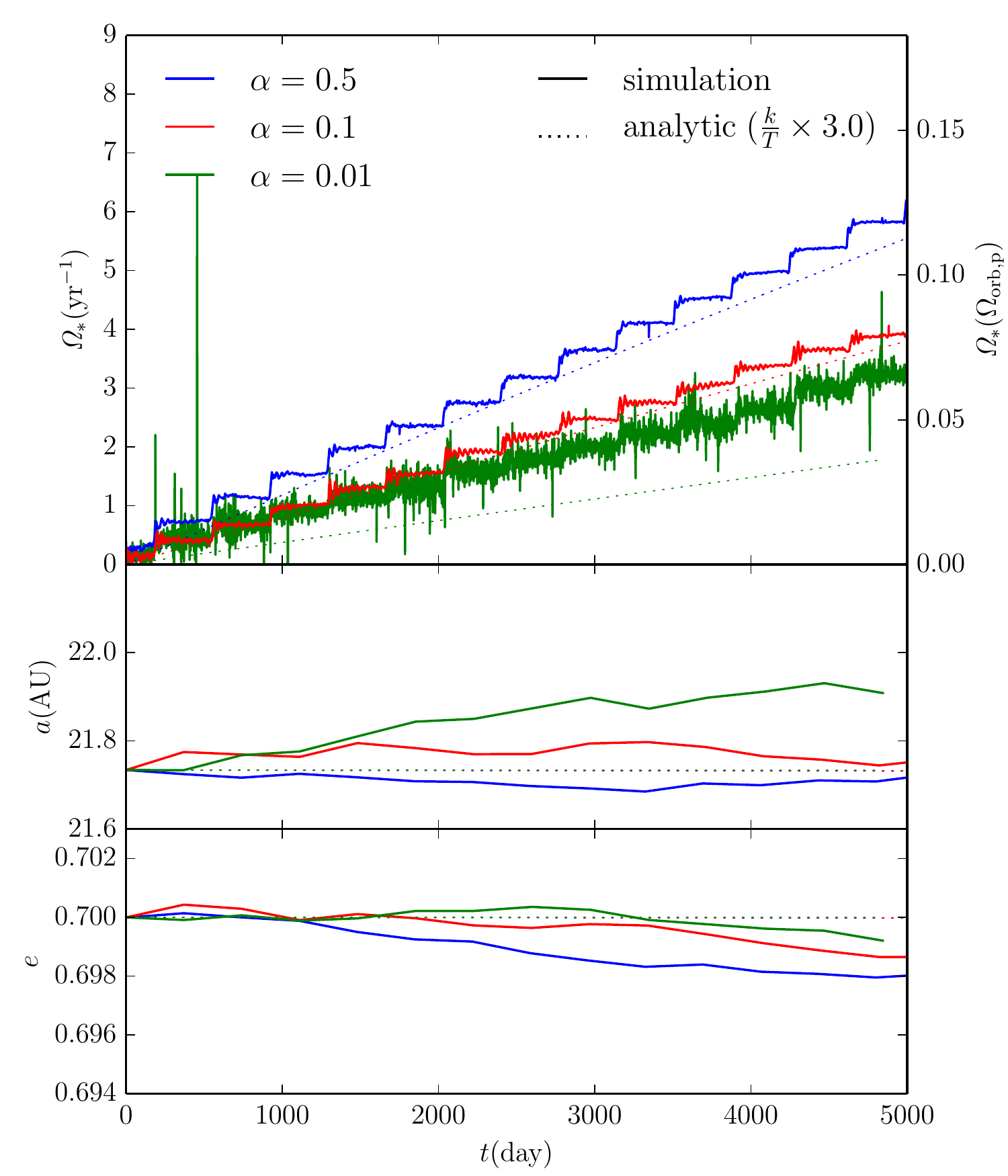}
            \caption{
                The long term evolution through tidal dissipation of $\AngV_*$, $a$ and $e$ for model O0R37 with varying values of $N$ and $\alpha$.
                Dotted lines show the analytical solution for tidal dissipation \citep{hut_tidal_1981} with $k/T \approx 3 (k/T)_\rasio$.
                Different values of $N$ and $\alpha$ result in a different radius of the relaxed SPH realization of the star, the analytical solution matching each simulation is therefore calculated with the effective radius of the SPH realization of the star in that simulation.
                The simulations with $N=100\,000$ and $N=200\,000$ are more computationally expensive and we have therefore decided to run them for only 2\,000\,days.
                Note that the blue line in both plots is the same simulation.
                \label{fig: long term rotation}}
        \end{figure*}

        Before investigating the combined effects of tidal dissipation and mass loss on the stellar orbit, we isolate the effect of tidal dissipation.
        We choose a simulation in which the stellar radius $R_* < R\roche$ (model O0R37, see table~\ref{tab: varied simulation parameters}) so that mass loss is negligible.
        In the absence of mass loss, the change in the orbital parameters is dominated by the effects of tidal dissipation.

        In figure~\ref{fig: long term rotation} we present the evolution of $\AngV_*$, $a$ and $e$ over a period of 5\,000\,days ($\unsim 13$\,orbits).
        All simulations use the initial conditions of model O0R37 while we varied $N$ (left figure) and $\alpha$ (right figure).
        It can be seen that for different values of these SPH parameters we find different values of $\dot{\AngV_*}$.

        The difference in $\dot{\AngV_*}$ can be understood by taking the radius of the star into account.
        Because of the quantization approximations inherent to SPH, a different value of $N$ can result in a slightly different hydrodynamical balance within the star.
        This difference, or even a different value of $\alpha$, can result in variation of the radius of the SPH realization of the star after relaxation.
        Even a small change in $R_*$ can account for the differences in $\dot{\AngV_*}$ because $\dot{\AngV_*} \propto R_*^6$ \citep{hut_tidal_1981}.

        While we do not include the physical processes that regulate tidal damping in our simulations, the artificial viscosity in SPH can cause dissipation.
        To interpret the results of our simulation, it is good to know how the timescale of this artificial dissipation compares with the timescale of realistic physical dissipation.
        The dotted line in figure~\ref{fig: long term rotation} is a numerical integration of the analytical solution for the change in orbital parameters through tidal dissipation.
        In this analytical solution, we have used the radius of the relaxed SPH realization of the star for each simulation.
        The analytical solution is close to the result of our simulation when we assume $k/T = 3 \times (k/T)_\rasio$, which is within the uncertainty in the tidal damping time-scale.
        This is fortuitous because we can now directly compare the tidal dissipation in our simulations with a real physical system.

        However, in the bottom panels of figure~\ref{fig: long term rotation} we see that the orbit circularizes ($e$ decreases) faster than what is predicted by the analytical models.
        We also see larger changes in $a$ than what is predicted by the analytical models, but there is no clear trend in these changes.
        We have carefully examined these simulations and we find that these changes in $a$ and $e$ result from errors in the total energy ($E\tot$) and angular momentum ($J\tot$) in the simulations.
        These errors are typically of order $\Delta E/E\tot \sim \Delta J/J\tot \sim 10^{-4}$ per 1200\,days, which is reasonable for a gravitational tree code with smoothing like the one in \ficode.
        However, in figure~\ref{fig: long term rotation} we see that these small errors have a larger effect on the orbital parameters than the tidal dissipation.
        We will therefore carefully examine the contribution of these errors when we investigate the orbital evolution in section~\ref{sec: Orbital evolution}.

    \subsection{Mass loss} \label{sec: Mass loss}
        \begin{figure*}
            \includegraphics[width=\textwidth]{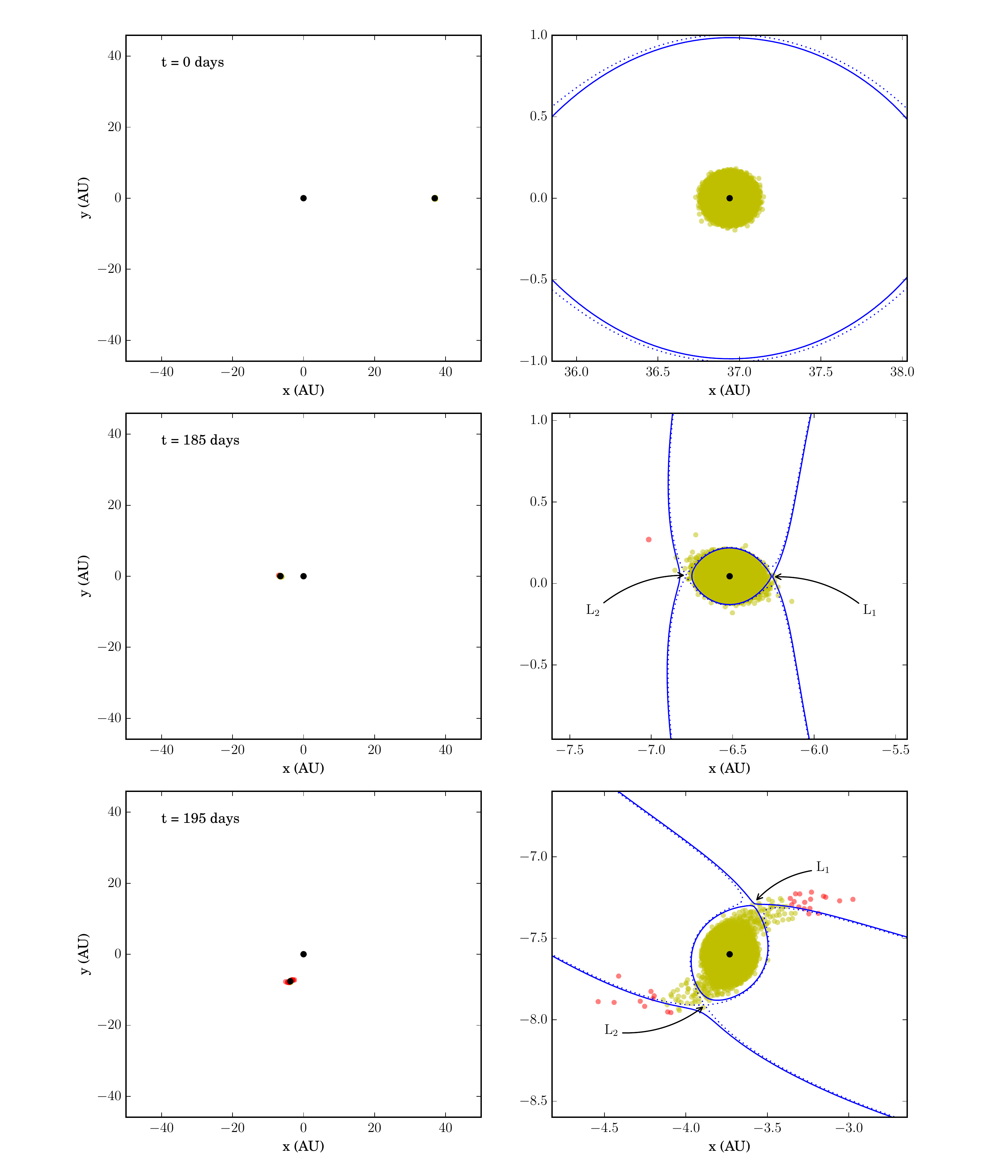}
            \caption{
                The particle positions in the orbital plane for three snapshots of the simulation with model O1R43.
                In the left panels we present an overview while the right panels zoom in on the star position.
                Black circles are the black hole and the stellar core, yellow and red circles are SPH particles bound and not bound to the star respectively.
                The blue solid and dotted lines are approximate equipotential surfaces through \Lone\ and \Ltwo\ respectively.
                The top panel shows the apocentre position and the middle panel shows the pericentre position.
                The bottom panel shows that after the pericentre passage, particles leave the star through both the \Lone\ and \Ltwo\ points.
                \label{fig: sph}}
        \end{figure*}

        \begin{figure}
            \includegraphics[width=0.5\textwidth]{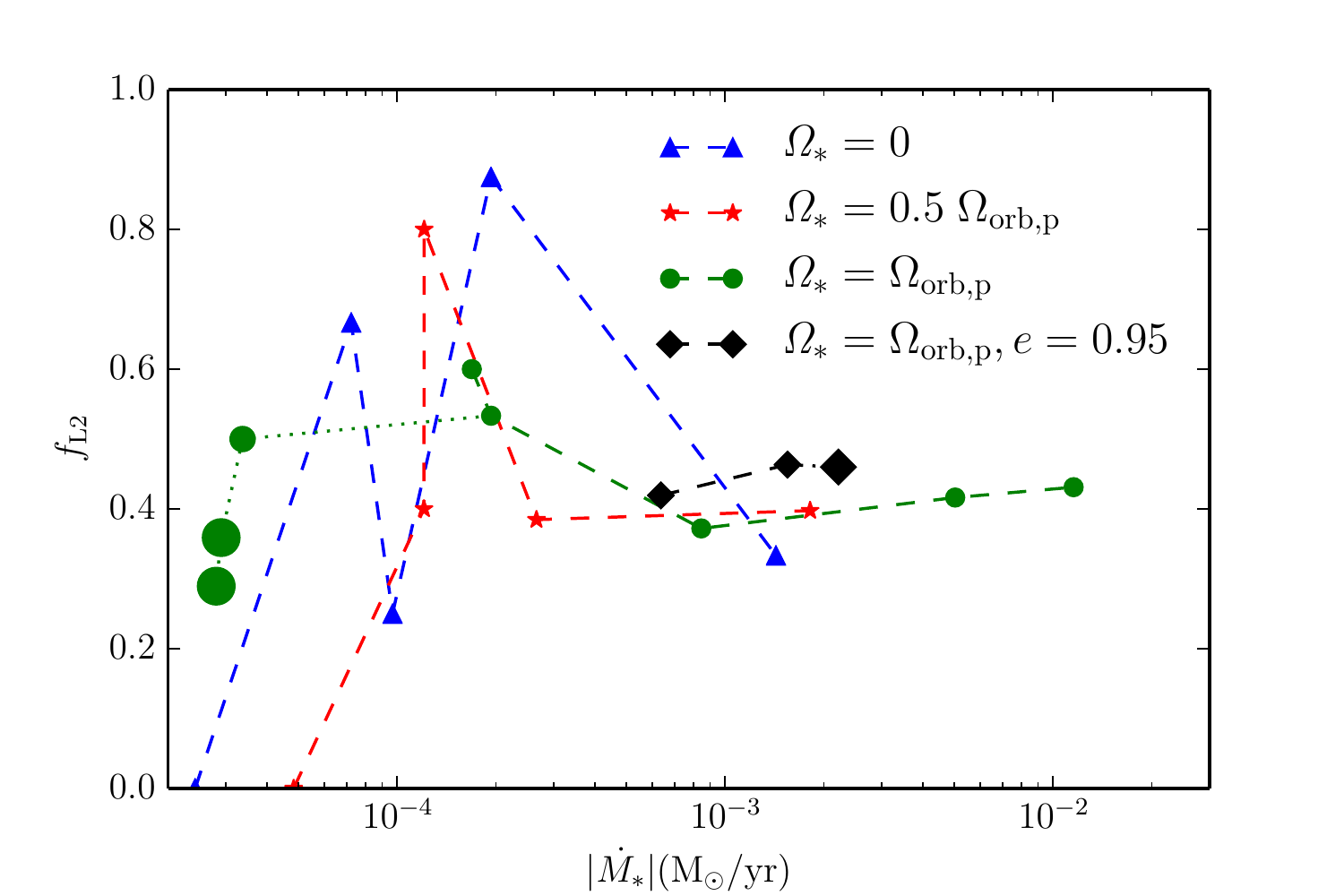}

            \caption{
                The fraction of mass lost through the \Ltwo\ point ($f_{\Ltwo}$) as opposed to the \Lone\ point as a function of the total mass loss rate.
                Sets of simulations with only $R_*$ differs between them are connected by dashed lines and larger symbols connected by dotted lines represent the higher resolution simulations listed in table~\ref{tab: varied simulation parameters}.
                For smaller total mass loss rate, there is a larger spread because only a small number of SPH particles are lost in this case.
                The general trend is that almost half the mass is lost through the \Ltwo\ point and moves away from the black hole.
                \label{fig: l2 mass loss}}
        \end{figure}

        \begin{figure}
            \includegraphics[width=0.5\textwidth]{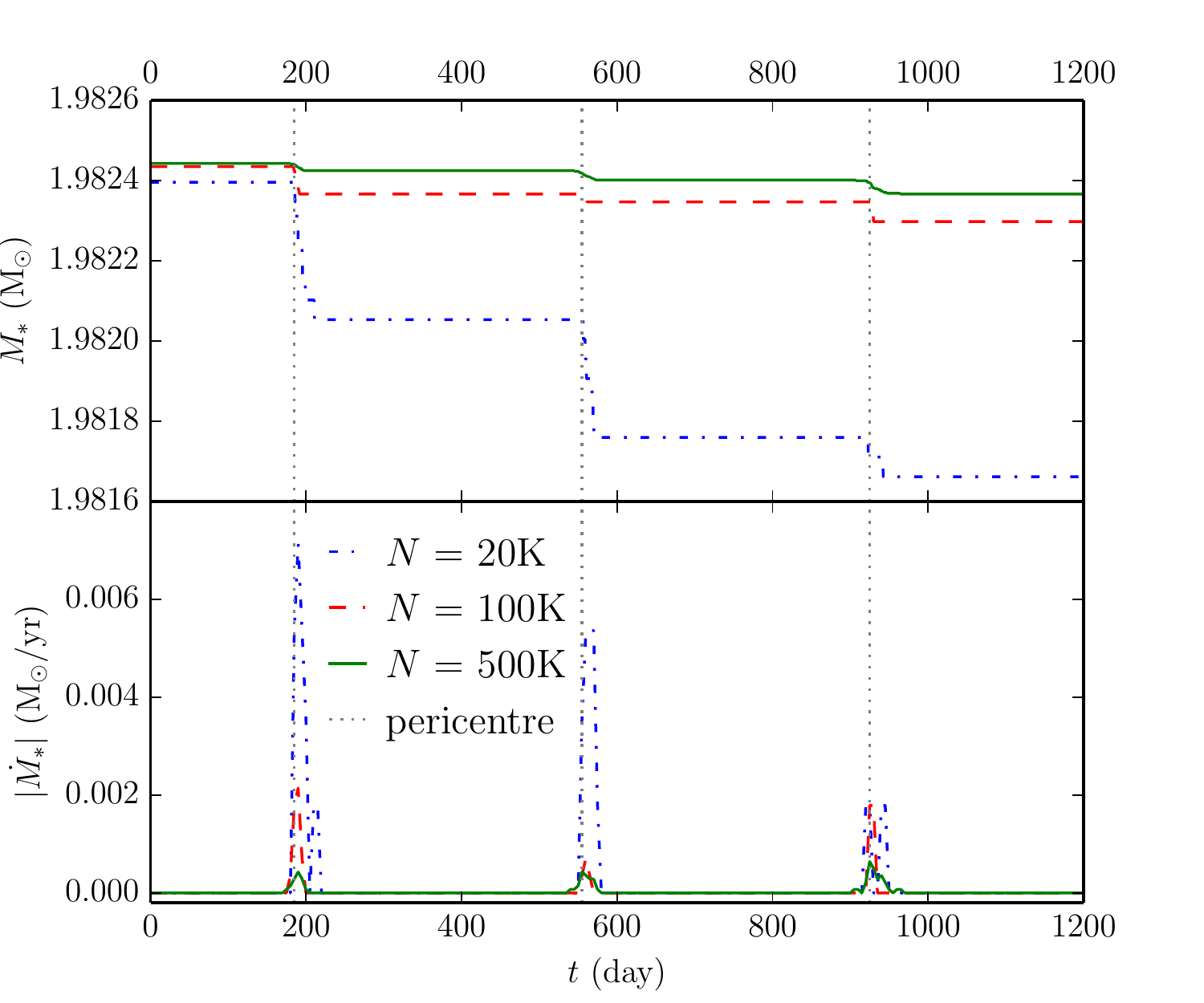}

            \caption{
                Total mass and mass loss rate of model O1R39 with $N =$ 20\,000, 100\,000 and 500\,000 particles.
                The mass loss rate is lower for simulations with a higher resolution.
                \label{fig: mass loss resolution}}
        \end{figure}

        \begin{figure}
            \includegraphics[width=0.5\textwidth]{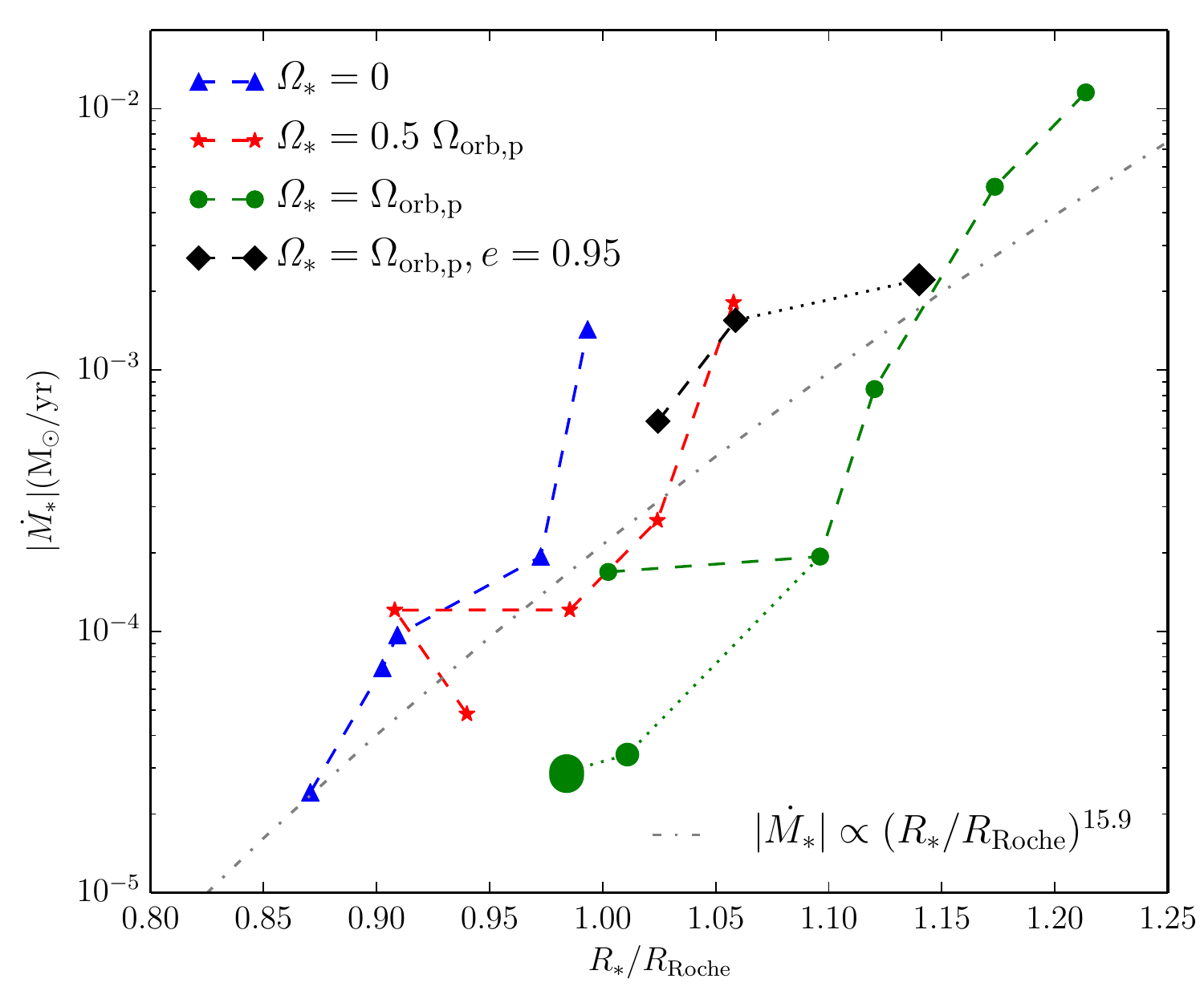}

            \caption{
                The average mass loss for different stellar radii (in units of the Roche radius).
                Only simulations with significant mass loss are included.
                The dash-dotted line is a fit to these points, but note that the mass loss is resolution dependent.
                \label{fig: radius mass loss}}
        \end{figure}

        \begin{figure}
            \includegraphics[width=0.5\textwidth]{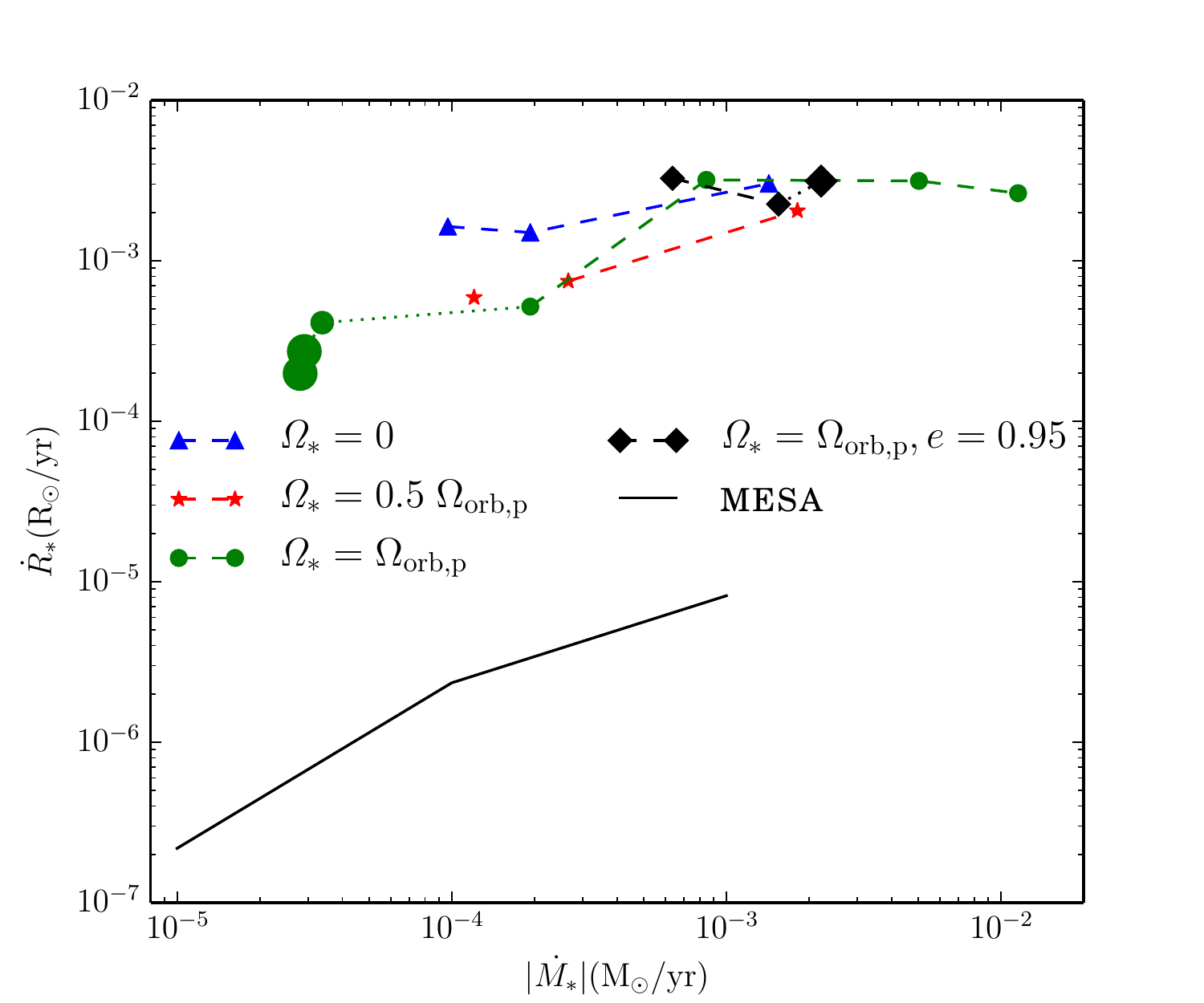}

            \caption{
                The expansion of the star in response to mass loss in our simulations compared to a stellar evolution model using \mesa\ with artificial mass loss.
                In our SPH simulations, $\dot{R_*}$ is over two orders of magnitude higher than in the \mesa\ model.
                \label{fig: rdot mass loss}}
        \end{figure}

        \begin{figure}
            \includegraphics[width=0.5\textwidth]{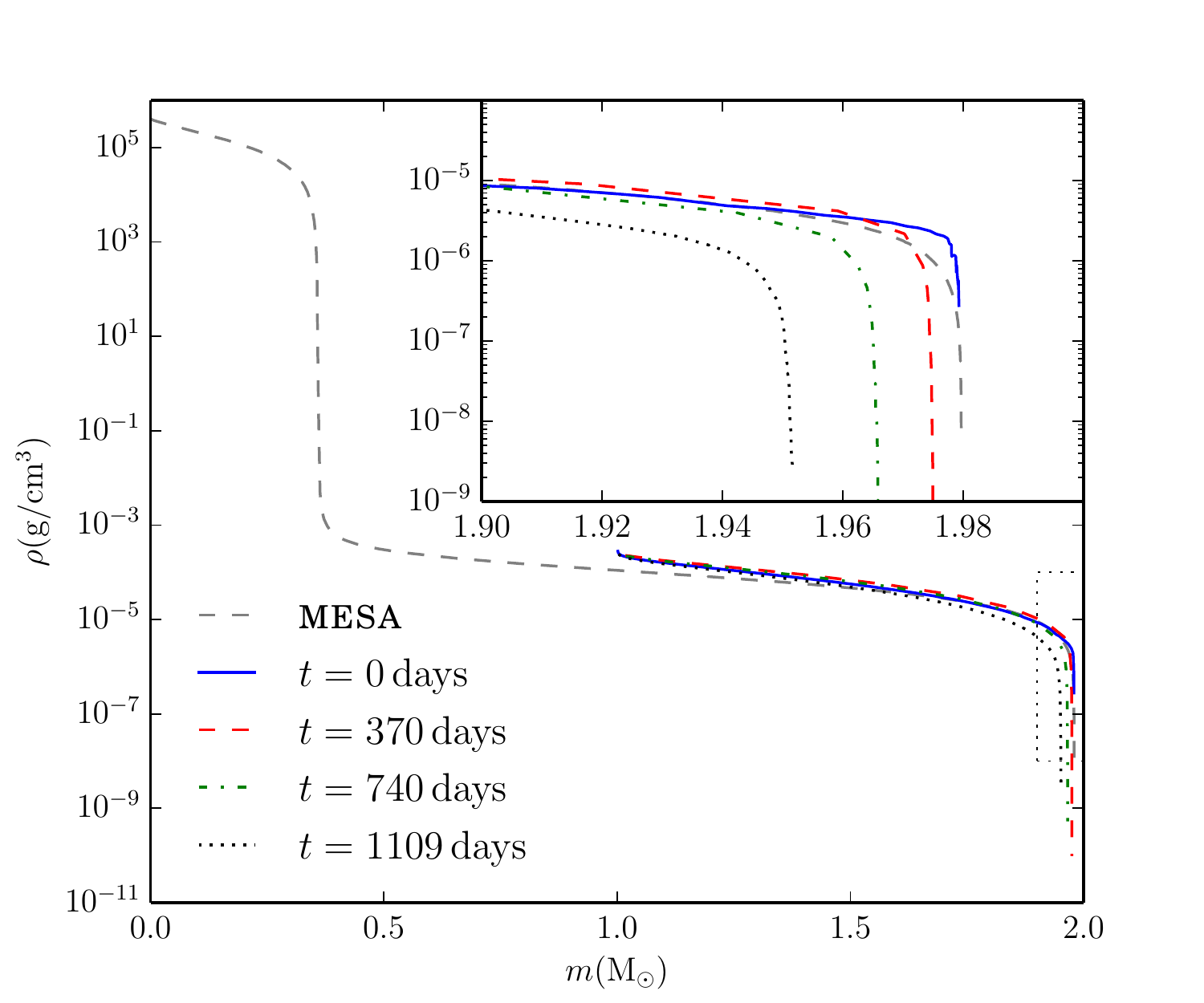}
            \caption{
                The stellar density profile for model O1R45 at the apocentre passages of the simulation, compared to the \mesa\ model.
                As mass is lost, the density at the edge of the star decreases, but the overall shape of the density profile remains the same.
                \label{fig: density time}}
        \end{figure}

        The star loses mass near pericentre in all simulations with $R_* \gtrsim R\roche$.
        We present three snapshots of a simulation with relatively high mass loss rate (O1R43) in figure~\ref{fig: sph} as an example.
        At pericentre ($t=185$ days) the star fills its Roche-lobe, but the mass is only lost after the pericentre passage.

        Part of the stellar mass is lost through the second Lagrangian point (\Ltwo), instead of the first Lagrangian point (\Lone), which can be seen in the bottom right panel of figure~\ref{fig: sph}.
        The binary system we study has a high mass ratio and therefore the difference in gravitational potential between the \Lone\ and \Ltwo\ points is small.
        If the stellar surface would follow an equipotential surface perfectly, this small difference in the gravitational potential would be enough to cause all mass to leave the star through the \Lone\ point.
        However, the stellar surface layers have a finite thickness, so mass loss through the \Ltwo\ point is a physical possibility.
        In figure~\ref{fig: l2 mass loss} we present the fraction of mass lost through the \Ltwo\ point ($f_\Ltwo$) in our simulations.
        In simulations where the total mass loss per orbit is large compared to the SPH particle mass, we find that $f_\Ltwo \approx 0.4$.
        For simulations with lower mass loss rate, there is a larger spread in $f_\Ltwo$, which is a result of the limited resolution in those simulations.

        In figure~\ref{fig: mass loss resolution} we present the stellar mass and mass loss rate as a function of time for model O1R39 with $N = 20\,000$, $N = 100\,000$ and $N = 500\,000$.
        The mass loss rate is smaller in simulations with higher resolution (larger $N$).
        There are two related numerical effects in our simulations that could explain this behaviour.
        The density at the outer edge of the star is lower in simulations with a higher resolution (figure~\ref{fig: radial density}), which allows less mass to escape the star.
        The mass loss rate depends on the radius of the star, and the effective radius of the relaxed SPH realization of the star differs between the simulations with different resolution.
        Because the resolution affects the mass loss rate, we will not draw any conclusions based on the absolute mass loss rate; instead we focus on the effect of a given mass loss rate on the orbital evolution.

        Using figure~\ref{fig: mass loss resolution} we confirm that the mass is mostly lost after pericentre.
        We find that there is a time delay of up to $10$\,days between the Roche-lobe overflow and the peak of the mass loss for systems where $e=0.7$.
        For systems where $e=0.95$, we do not find any delay larger than our time resolution limit of 1\,day.
        When the gas is gravitationally accelerated during the pericentre passage, it takes some time before it becomes unbound.
        We suspect that this time delay is related to the hydrostatic timescale (\thydr) of the star \citep{kippenhahn_stellar_2012}.
        Indeed for the star used in simulations with $e=0.7$, $\thydr \approx 2$\,days while for simulations with $e=0.95$, the smaller star has $\thydr \approx 0.2$\,days.

        In figure~\ref{fig: radius mass loss} we present the mass loss rate as a function of initial radius as calculated in section~\ref{sec: stellar radius}.
        We find that the mass loss rate is extremely sensitive to the radius ($|\dot{M_*}| \propto (R_*/R\roche)^{15.9}$).
        In fact, most of the dependence of the mass loss rate on the resolution seen in figure~\ref{fig: mass loss resolution} can be accounted for by the difference in the effective radius.
        We find a spread of more than an order of magnitude around the $\dot{M_*}$-$R_*$ relationship which could be partly due to numerical noise.

        In figure~\ref{fig: rdot mass loss} we present the rate of change in the stellar radius ($\dot{R_*}$) caused by the stellar mass loss.
        Losing mass from the surface of the star results in a lower pressure at the stellar surface.
        This low pressure will cause the inner part of the SPH model to respond by expanding and therefore the stellar radius increases during mass loss.
        This is also true, at least qualitatively, for a realistic star with a convective envelope, and therefore we have also plotted $\dot{R_*}$ calculated with the stellar evolution code (\mesa) where we have set an artificial mass loss for the case in which $R_* = 39\,\RSun$.
        However, quantitatively $\dot{R_*}$ is not the same because we do not have a realistic stellar structure.
        Our simulations do not include nuclear burning in the stellar core and in fact we have replaced the entire core with a single SPH particle.
        More importantly, we cannot fully resolve the steep density gradient at the edge of the star using SPH.
        The expansion of the star may also be affected by the gravitational interaction with the black hole.
        During this interaction, orbital energy can by transferred to heat, which can cause the star to expand.
        This process may be partly responsible for the stellar expansion in the SPH model, while it is not taken into account in the \mesa\ model.
        We show in figure~\ref{fig: rdot mass loss} that the change in the radius of our SPH star is more than two orders of magnitude higher than what is calculated using \mesa.
        It is unclear which part of this difference is caused by the lack of a realistic stellar structure in our simulations, and which part is caused by the more realistic effect of the gravitational interaction.
        We therefore conclude that we cannot ensure that we can reliably simulate the system with mass loss for a long period of time.
        Even during the 1200\,days used in these simulations, we already see a slight increase in the mass loss rate caused by the increase in the stellar radius.

        In figure~\ref{fig: density time} we present the stellar density profile during different apocentre passages in the simulation.
        The star loses mass from the outer region so mass loss causes the density at the edge of the star to drop.
        Later in the simulation, the density further from the edge of the star also becomes lower as the stellar structure adjusts to the mass loss.
        Despite the mass loss, the general shape of the stellar density profile remains the same.
        This indicates that the time that the star is far from the black hole is long enough for the stellar structure to adjust to the mass loss.

    \subsection{Orbital evolution} \label{sec: Orbital evolution}
        \begin{figure}
            \includegraphics[width=0.5\textwidth]{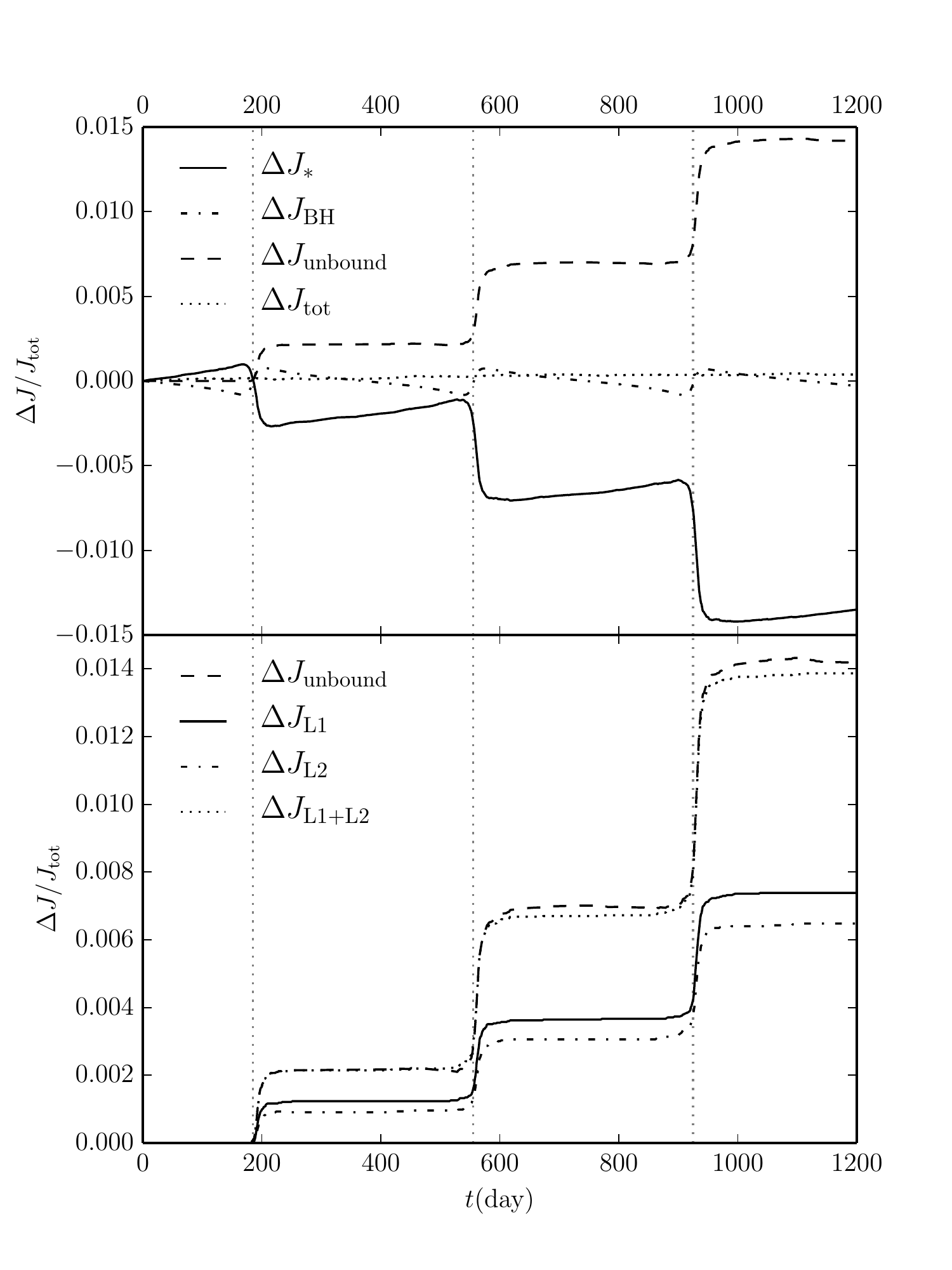}
            \caption{
                We present the change in angular momentum in the orbital plane of the star, black hole and unbound gas for model O1R45 (top panel).
                The star and black hole exchange angular momentum because the orbit is eccentric and the star loses angular momentum that is taken by the mass that is lost.
                The total angular momentum is conserved, with an error of $\unsim 10^{-4}\,J\tot$.
                We also present details on different contributions to the angular momentum in the unbound gas (bottom panel).
                We compare the angular momentum taken by the mass as it is lost through \Lone\ and \Ltwo\ ($\Delta J_\Lone + \Delta J_\Ltwo = \Delta J_{\Lone + \Ltwo}$) with the total angular momentum of the unbound gas ($\Delta J\unbound$)
                \label{fig: angular momentum}}
        \end{figure}

        \begin{figure}
            \includegraphics[width=0.5\textwidth]{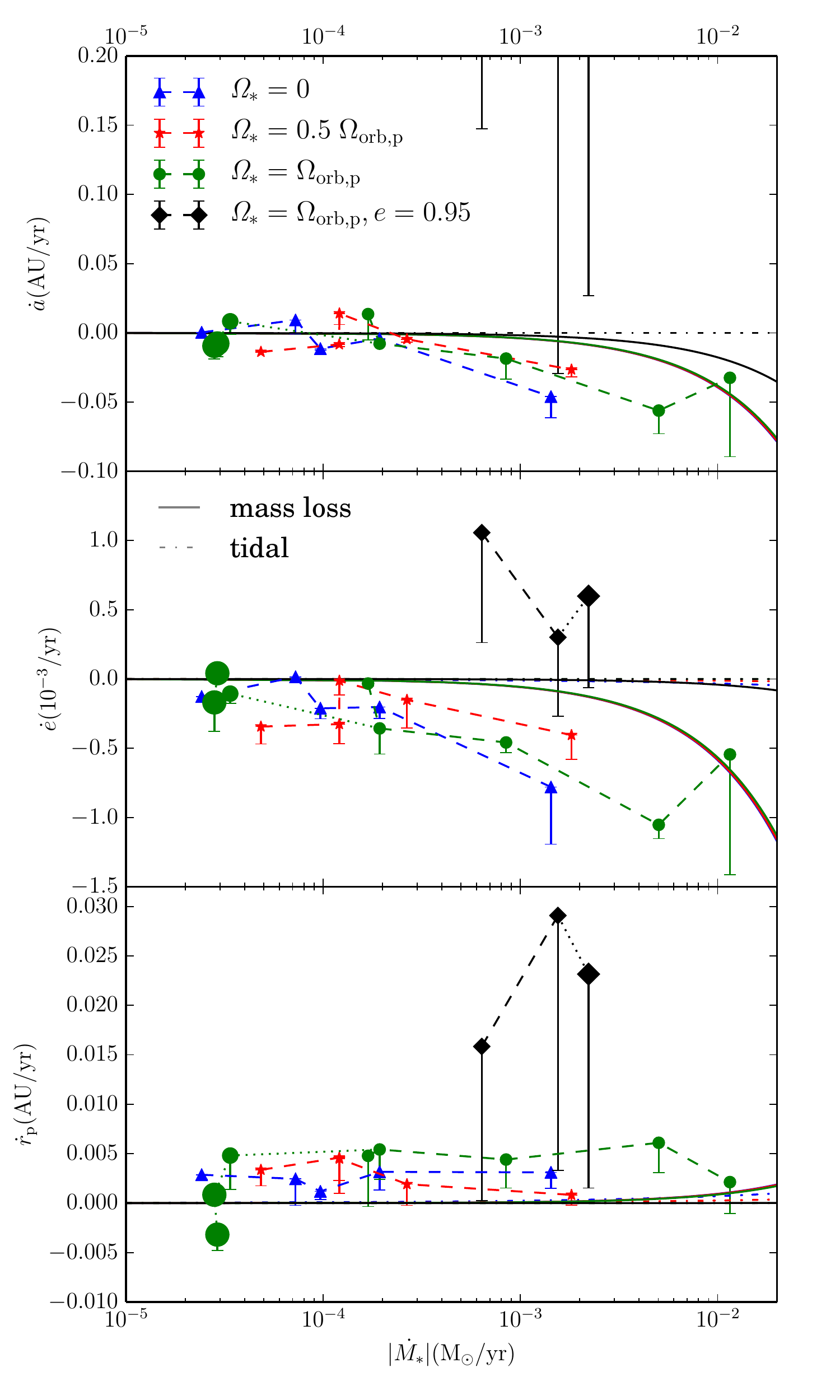}

            \caption{
                The change in orbital parameters as a function of mass loss rate for our simulations, $e=0.7$ unless otherwise specified.
                The error bars are based on the systematic errors in $E\tot$ and $J\tot$, the details of these are discussed in the text.
                The solid and dash-dotted lines represent the analytical predictions for mass loss \citep{sepinsky_interacting_2009} and tidal dissipation \citep{hut_tidal_1981} respectively.
                For the analytical prediction from mass loss, we assume that the mass lost through $L_2$ is not accreted onto the black hole.
                The tidal effects depend on the mass loss rate because we have varied the radius with the mass loss rate following the fit in figure~\ref{fig: radius mass loss}.
                In the top panel, the simulations with $e=0.95$ have $\dot{a} \sim 0.9$\,AU/yr but due to the large errors the result are consistent with zero.
                \label{fig: combined evolution}}
        \end{figure}

        The mass that leaves the star has angular momentum, which is removed from the star and therefore the stellar orbit changes.
        In figure~\ref{fig: angular momentum} (top panel) we present the change in the angular momentum of the star ($\Delta J_*$), the black hole ($\Delta J\BH$), and the matter that is not bound to the star ($\Delta J\unbound$).
        The star and the black hole exchange angular momentum periodically during the eccentric orbit while the star loses angular momentum due to the mass loss.
        The error in the total angular momentum $\Delta J/J\tot \sim 10^{-4}$ just like in the simulations without mass loss, but the effect of mass loss on $J_*$ is larger than the effect of this error.

        In figure~\ref{fig: angular momentum} (bottom panel) we distinguish between the angular momentum taken by matter leaving the star through \Lone\ ($\Delta J_\Lone$) and through \Ltwo\ ($\Delta J_\Ltwo$).
        The matter leaves the star in opposite directions, but both remove a positive amount of angular momentum from the star, and therefore the effect on the star is cumulative.
        All the angular momentum taken when matter leaves the star ($\Delta J_{\mathrm{\Lone + \Ltwo}}$) is nearly equal to the angular momentum in the unbound matter ($\Delta J\unbound$).
        We therefore conclude that the gravitational interaction after mass loss between the star and the unbound matter is negligible.

        In figure~\ref{fig: combined evolution} we present the change in the orbital parameters $a$, $e$ and pericentre distance (\rp) of our simulations as a function of mass loss and compare these with analytical models for mass loss and tidal dissipation.
        As noted in section~\ref{sec: Tidal dissipation} we have to include the effect of the errors in energy and angular momentum.
        We propagate these errors but they appear to be systematic and always in the same direction, we therefore only show the error bars in that direction.
        Note that correct treatment of these systematic errors and further statistical analysis would require knowing the correct model of non-linearity \citep[e.g.][]{barlow_asymmetric_2004}.
        Since our knowledge about the nature of these errors is not sufficient for such complex treatment, we must be very cautious when interpreting these results.

        In the top two panels of figure~\ref{fig: combined evolution} we see that $a$ and $e$ generally decrease for simulations with $e=0.7$, and that the rate of change is larger for larger mass loss, which is qualitatively consistent with the analytical models.
        In the bottom panel we see that \rp\ generally increases with time at a higher rate than what is predicted by the analytical models but they appear to be consistent within the error bars.
        For the simulations with $e=0.95$, the errors are far larger because of the higher velocity and acceleration at pericentre that amplify the errors.
        We therefore cannot draw any conclusions about the orbital evolution of these systems.

        With the higher value of \rpdot\ we have measured, we can calculate the effective time-scale for the orbital evolution in our simulations, which is $\unsim 10^3$\,years.
        We can then compare that with the time-scale of the change in $R_*$ through stellar evolution with mass loss from figure~\ref{fig: rdot mass loss}, which is $\unsim 10^6$\,years.
        The change in the stellar orbit is clearly much faster, and dominates the evolution of this system.

    \subsection{Comparison with observation}
        \begin{figure}
            \includegraphics[width=0.5\textwidth]{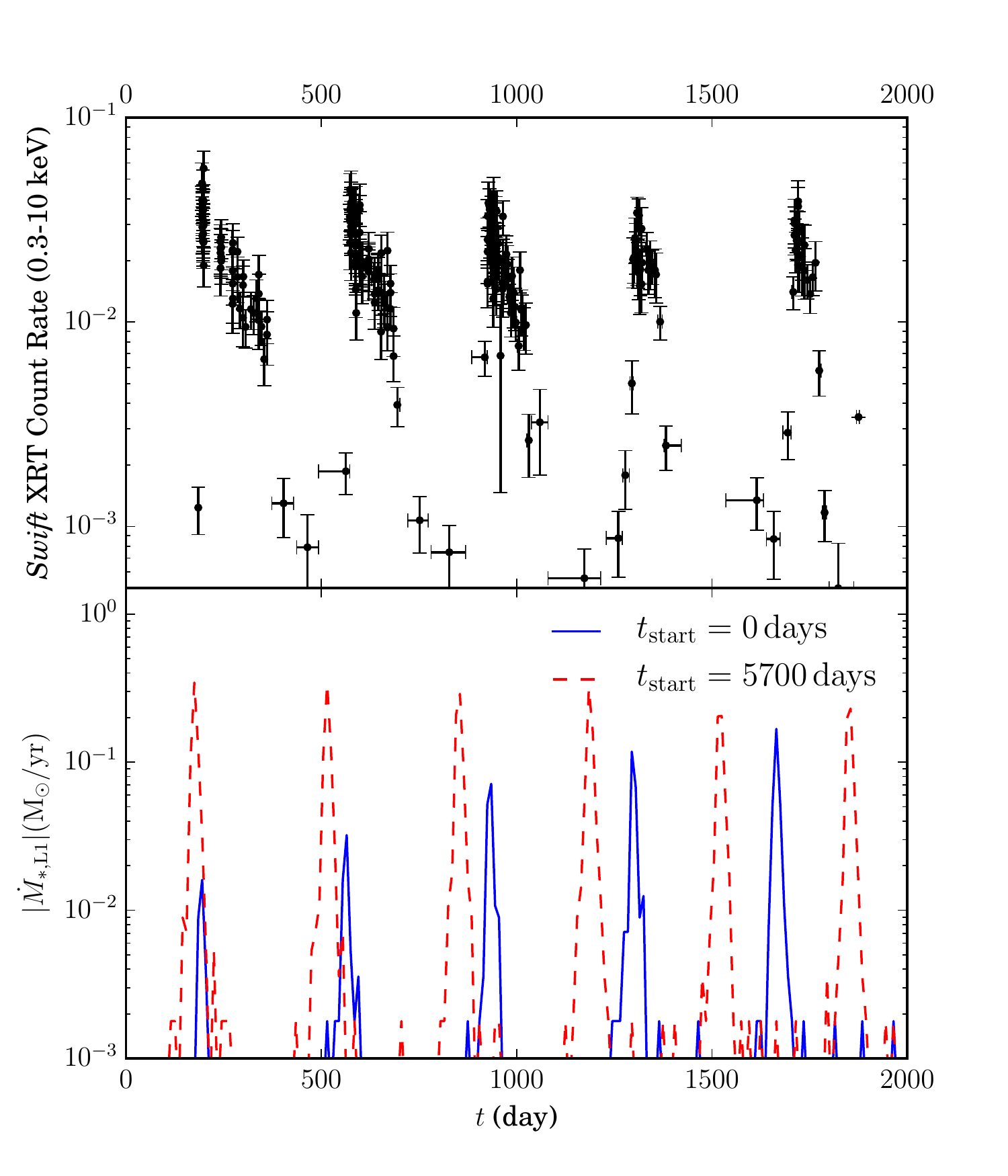}
            \caption{
                The observed X-ray light curve from \textit{Swift} \citep[top,][]{godet_implications_2014}, compared with the mass loss through \Lone\ from two parts of simulation O1R43 (bottom). The start time is chosen to provide the closest match between the observations and the simulations.}
                \label{fig: flux}
        \end{figure}

        In figure~\ref{fig: flux} we compare the observed X-ray light curve with the mass loss rate through \Ltwo\ as a function of time in simulation O1R43.
        We cannot compare the mass loss rate with the light curve directly because several processes, like the formation of an accretion disc, will modulate the signal.
        This accretion disc, and in particular the viscous time-scale of that disc, are an important and heavily debated part for most models of HLX-1 \citep[see e.g.][]{soria_eccentricity_2013}.
        We cannot draw conclusions about the accretion disc from our simulations, however, we can compare our results with the observations if we make two assumptions:
        1) The X-rays are caused by the mass that is lost from the star and the efficiency with which the mass is turned into X-rays does not differ noticeably between outbursts.
        2) Any delay between the time of mass loss and the time when X-rays are emitted does not differ between outburst.
        It should be noted that accretion discs physics is very complex and therefore these assumptions do not necessarily hold for all accretion scenarios.

        At the start of the simulation ($t_\mathrm{start} = 0$\,days), the mass loss rate increases, which does not match the observations.
        After running the simulation for a longer time ($t_\mathrm{start} = 5700$\,days), the mass loss rate starts to decrease, matching the observations, although the orbital period has changed by that time.
        Note however, that the long term mass loss rate evolution is also influenced by numerical effects as discussed in section~\ref{sec: Mass loss}.

        The observed X-ray flux profile has a tail after every outburst while the simulated mass loss rate only shows a sharp peak at each outburst.
        The observed tail is probably caused by an accretion disc that can delay the accretion of material onto the black hole.
        Based on this assumption, \citet{soria_eccentricity_2013} have calculated that the accretion disc has a radius of $\unsim 15$\,\RSun.
        This is an order of magnitude lower then the radius estimate based on the observed continuum emission from the hot disc.
        This discrepancy is still an unsolved issue in most models for \hlx\ and our simulations are not able to resolve that.

        It is important to note that our simulations cannot explain the observed delay of over 30\,days and 60\,days in the last two outburst.
        Despite the fast evolution of the orbit, and particularly \rp, such strong variations do not occur in our simulations.

\section{Discussion and Conclusion}
    The analytical solutions that have been used so far to investigate the orbital evolution of eccentric binaries during mass transfer are not sufficient to predict the orbital evolution of the \hlx\ system.
    Key assumptions in the analytical mass loss model, that all mass is lost through the \Lone\ point at pericentre and that the mass transfer happens instantaneously, turn out to be incorrect.
    The orbital evolution in our simulations is faster than what is predicted by analytical models but within the errors they are still consistent.
    However, even with full hydrodynamical mass transfer simulations we are unable to explain the observed delay of over 30\,days in the last outburst.

    The density near the edge of the star, and therefore $\dot{M}_*$, depends on the number of SPH particles used in our simulation (figures~\ref{fig: radial density}~and~\ref{fig: mass loss resolution}).
    We are therefore not able to investigate the mass loss rate as a function of stellar radius ($\dot{M}_*(R_*)$).
    However, the goal of this research is to investigate the orbital evolution of \hlx, not the mass loss rate of the star.
    To account for the uncertainty in $\dot{M}_*(R_*)$, we only investigate the change in orbital parameters as a function of $\dot{M}_*$, not as a function of $R_*$.
    We also note that the time-scale of tidal damping is within the range of theoretical expectations, although this is probably coincidental as the detailed physics of tidal dissipation is not accounted for in our work (section~\ref{sec: Tidal dissipation}).

    We have shown that the error in angular momentum (and energy) is small ($\Delta J/J\tot \sim 10^{-4}$, figure~\ref{fig: angular momentum}), but that this still has a considerable effect on the orbit.
    The change in the orbital angular momentum due to this error is larger than the change due to tidal dissipation, but smaller than the change due to mass loss.
    Therefore, we have reason to believe that apart from these errors, our simulations provide a reliable prediction for the orbital evolution of \hlx.
    We obtain the following results:

    \begin{description}
        \item[$\bullet$]
        Approximately 40\% of the stellar mass is lost through the \Ltwo\ point instead of through the \Lone\ point and this mass carries approximately 40\% of the angular momentum that is lost from the star.
        This agrees with the results for mass transfer in eccentric binaries \citep{regos_mass_2005,lajoie_mass_2011}.
        The analytical models for binary evolution could be improved by taking this into account.
        The angular momentum carried by the mass lost through the \Lone\ point and the \Ltwo\ point have the same sign.
        Therefore, the effect of the mass loss through \Lone\ and \Ltwo\ on the stellar orbit is cumulative.

        \item[$\bullet$]
        Most of the mass is lost up to 10\,days past the pericentre passage.
        The timescale of this delay appears to be related to the hydrostatic timescale of the star, which was also noted by \citet{lajoie_mass_2011}.
        In an eccentric orbit, the star has a different velocity and a different distance to the accretor due to this delay.
        The angular momentum carried away by the mass loss is therefore also affected by this delay and it should be taken into account in predictions of the orbital evolution of the system.
        Taking this delay into account would therefore also improve the analytical models.

        \item[$\bullet$]
        When $e=0.7$, we find that the semimajor axis ($a$) and eccentricity ($e$) generally decrease, and this decrease is faster for higher mass loss rates, which is qualitatively consistent with analytical models.
        In nearly all models, we find that the pericentre distance (\rp) increases faster than what is predicted by analytical models, but within the error they are still consistent.

        \item[$\bullet$]
        When $e=0.95$, the behaviour of our simulations is very different from what is predicted by analytical models.
        However, the errors in energy and angular momentum become so large that we cannot draw any conclusions regarding the loss of angular momentum.

        \item[$\bullet$]
        We cannot explain the observed delay in the outbursts of \hlx\ as a change in the orbit seen in our simulations, despite the fast changes in the orbit that we do see.

    \end{description}

    If we take the increase in \rp\ that we have found in our simulations at face value, we can use this to constrain the possible formation mechanisms of the \hlx\ system.
    An increase in \rp\ can cause a decrease in the mass loss rate if no other processes would affect the mass loss rate.
    At the start of our simulations we instead see an increase of the mass loss rate but this is most likely caused by the unphysical expansion of the star.
    If the mass loss rate would indeed decrease, then this would qualitatively agree with the observation that the total outburst fluence decreases.
    However, the value of \rpdot\ we find is very high and therefore the mass loss rate would decreases very quickly under this assumption.
    Extrapolating back in time, the mass loss should have started within the last $\unsim 10$\,years to avoid unphysically high mass loss rates in the past if the value of \rpdot\ we measured is correct.
    This conclusion agrees with the fact that \hlx\ was not detected in ROSAT observations in the early nineties \citep{webb_chandra_2010}.
    The change in radius through stellar evolution is too slow to have initiated the mass transfer.
    However, we cannot exclude the possibility that a change in the stellar rotation, caused by tidal dissipation, could have initiated the mass transfer.

    If the star did not evolve on the current orbit, it could have been captured by the black hole within the last $\unsim 10$\,years.
    This could have happened through tidal capture as discussed below.
    Another possibility is that a binary was disrupted by the black hole, leaving one star on an eccentric orbit, while the other star escaped the system.
    The probabilities for both these scenarios are largely unknown \citep[but see][]{caputo_estimating_inprep}.
    Recently, alternative scenarios have also been proposed, such as wind accretion \citep{miller_wind_2014} and extreme super-Eddington accretion on a stellar mass object \citep{king_hlx-1_2014}.

    We can compare our results with the results from the work by \citet{godet_implications_2014} as our investigation and methods are similar.
    We should take into account that we have used very different initial conditions and therefore have investigated a different scenario.
    The most important difference in our results is that we do not see large stochastic variations in the orbital period, which could simply be due to the far lower eccentricity used in this work.
    However, it could also be a result of the artificial viscosity; we have used $\alpha=0.5$, while \citet{godet_implications_2014} did not use artificial viscosity at all.
    This choice was made by them because artificial viscosity might lead to unphysical damping of stellar oscillations \citep{lombardi_tests_1999}.
    We note however, that viscous processes in real stars are only partially understood.
    This different choice for the artificial viscosity is also very important because it affects the tidal dissipation as we have discussed above.
    They do not discuss the possibly unphysical radius expansion we have noted in section \ref{sec: Mass loss}, although we suspect that it also affects the long term evolution of $\dot{M_*}$ in their simulation.
    The effects of the energy and angular momentum errors are also important to note, because we have found that these errors are far larger when the eccentricity is higher, and their model has an even higher eccentricity than what we have investigated.
    However, they calculate gravity with direct summation and find $\Delta E/E\tot \lesssim 10^{-4}$ and $\Delta J/J\tot \lesssim 10^{-7}$ which cannot explain the orbital evolution they find (J. Lombardi, private communications).

    Since the mass of the donor star is not constrained by observations, we have also experimented with $M\zams = 20\,\MSun$.
    We have tried this with both $e=0.7$ and $e=0.95$ and with $R_*$ determined using figure~\ref{fig: stellar radius}.
    However, these simulations were plagued by numerical instabilities.
    The difficulty we experienced in acquiring a dynamically stable solution may reflect a problem in the numerical methods.
    The high eccentricity in these orbits makes it very hard to integrate the equations of motion properly.
    In one of these cases we were able to acquire a numerically stable solution, but in that case the evolution of $a$ and $e$ behaved very unexpectedly and we were unable to determine if this was due to numerical issues or because such a system is intrinsically unstable.
    We have therefore omitted these results from this study.

    We have also attempted to extend the evolution of the system beyond three orbits in order to study the long term evolution of the mass transfer process.
    However, as we show in figure~\ref{fig: rdot mass loss}, the stellar radius does not evolve realistically using the SPH method.
    This raises the question if the SPH method is at all suitable to follow more than a single pericentre passage properly.
    In our simulations $\dot{R_*} \lesssim 4\times 10^{-3}$\,\RSun\,yr$^{-1}$, which corresponds to an 0.003\% increase during the 1200\,days over which we run our simulations.
    Since $|\dot{M_*}| \propto (R_*/R\roche)^{15.5}$, this radius increase is responsible for a 0.5\% increase in the mass loss rate.
    This is still a moderate error, but if we were to run for a much longer time, the increase in mass loss would accelerate the stellar radius expansion, resulting in a runaway process and this error would begin to dominate.
    We have therefore decided not to run our simulation for longer than 1200\,days in order to prevent the numerical method from directly influencing the measured mass loss rate in the simulations.
    Realizing this numerical effect, we have not interpreted the change in the mass loss rate in our comparisons with the observations.

    To further improve our understanding of the orbital evolution of \hlx, more simulations are required that address the caveats discussed in this work.
    Many of the initial parameters are unknown and we have only investigated a small subset of the possible values.
    A much larger set of these simulations, while computationally expensive, could reveal trends in the orbital evolution, and provide further understanding of the mechanisms involved.
    However, the large uncertainties in both the initial parameters and the accretion scenario in general as well as the numerical errors, should be addressed first.
    One way to address the numerical errors in energy and angular momentum would be to replace the gravitational tree method with a direct N-body method.
    In a tree method, the gravitational force from sets of distant particles is approximated as the force from a single mass.
    In a direct N-body method, the gravitational force from each particle is calculated separately.
    While more computationally expensive, this would improve the accuracy of the method and solve some of the problems encountered in this work \citep[for a comparison, see e.g.][]{bedorf_sparse_2012}.

    In conclusion, detailed simulations are crucial for investigating the dynamics of mass transfer.
    Analytical solutions are useful for a better understanding of the processes involved, but they can be improved upon using results from detailed simulations.
    We have been unable to explain the observed delay in the outbursts of \hlx\ with either the analytical models, or the simulations.
    It therefore seems unlikely that the periodicity in the outbursts corresponds to the period of an orbit as it is investigated here.

\section*{Acknowledgements}
    We thank the anonymous referee for critical reading that helped us to improve the manuscript.
    This work was supported by the Netherlands Research Council NWO (grants \#643.200.503, \#639.073.803 and \#614.061.608) and by the Netherlands Research School for Astronomy (NOVA).

\bibliography{HLX_1_paper_refs}

\begin{thebibliography}{}

\bibitem[\protect\citeauthoryear{Barlow}{Barlow}{2004}]{barlow_asymmetric_2004}
Barlow R.,  2004, in {PHYSTAT}2003, {SLAC} Asymmetric {Errors}.
ArXiv Physics e-prints

\bibitem[\protect\citeauthoryear{Bate, Bonnell \& Price}{Bate
  et~al.}{1995}]{bate_modelling_1995}
Bate M.~R.,  Bonnell I.~A.,    Price N.~M.,  1995, MNRAS, 277, 362

\bibitem[\protect\citeauthoryear{Baumgardt, Hopman, Portegies~Zwart \&
  Makino}{Baumgardt et~al.}{2006}]{baumgardt_tidal_2006}
Baumgardt H.,  Hopman C.,  Portegies~Zwart S.,    Makino J.,  2006, MNRAS, 372,
  467

\bibitem[\protect\citeauthoryear{B\'{e}dorf, Gaburov \&
  Portegies~Zwart}{B\'{e}dorf et~al.}{2012}]{bedorf_sparse_2012}
B\'{e}dorf J.,  Gaburov E.,    Portegies~Zwart S.,  2012, Journal of
  Computational Physics, 231, 2825

\bibitem[\protect\citeauthoryear{Belczynski, Kalogera, Rasio, Taam, Zezas,
  Bulik, Maccarone \& Ivanova}{Belczynski
  et~al.}{2008}]{belczynski_compact_2008}
Belczynski K.,  Kalogera V.,  Rasio F.~A.,  Taam R.~E.,  Zezas A.,  Bulik T.,
  Maccarone T.~J.,    Ivanova N.,  2008, ApJS, 174, 223

\bibitem[\protect\citeauthoryear{Caputo, de Vries, Patruno \&
  Portegies~Zwart}{Caputo et~al.}{prep}]{caputo_estimating_inprep}
Caputo D.~P.,  de Vries N.,  Patruno A.,    Portegies~Zwart S.,  in prep.

\bibitem[\protect\citeauthoryear{Church, Dischler, Davies, Tout, Adams \&
  Beer}{Church et~al.}{2009}]{church_mass_2009}
Church R.~P.,  Dischler J.,  Davies M.~B.,  Tout C.~A.,  Adams T.,    Beer
  M.~E.,  2009, MNRAS, 395, 1127

\bibitem[\protect\citeauthoryear{Davis, Narayan, Zhu, Barret, Farrell, Godet,
  Servillat \& Webb}{Davis et~al.}{2011}]{davis_cool_2011}
Davis S.~W.,  Narayan R.,  Zhu Y.,  Barret D.,  Farrell S.~A.,  Godet O.,
  Servillat M.,    Webb N.~A.,  2011, ApJ, 734, 111

\bibitem[\protect\citeauthoryear{de Vries, Portegies~Zwart \&
  Figueira}{de~Vries et~al.}{2013}]{de_vries_evolution_2013}
de Vries N.,  Portegies~Zwart S.,    Figueira J.,  2013, ArXiv e-prints

\bibitem[\protect\citeauthoryear{Eggleton}{Eggleton}{1983}]{eggleton_approximations_1983}
Eggleton P.~P.,  1983, ApJ, 268, 368

\bibitem[\protect\citeauthoryear{Eggleton, Kiseleva \& Hut}{Eggleton
  et~al.}{1998}]{eggleton_equilibrium_1998}
Eggleton P.~P.,  Kiseleva L.~G.,    Hut P.,  1998, ApJ, 499, 853

\bibitem[\protect\citeauthoryear{Farrell, Servillat, Gladstone, Webb, Soria,
  Maccarone, Wiersema, Hau, Pforr, Hakala, Knigge, Barret, Maraston \&
  Kong}{Farrell et~al.}{2014}]{farrell_combined_2014}
Farrell S.~A.,  Servillat M.,  Gladstone J.~C.,  Webb N.~A.,  Soria R.,
  Maccarone T.~J.,  Wiersema K.,  Hau G. K.~T.,  Pforr J.,  Hakala P.~J.,
  Knigge C.,  Barret D.,  Maraston C.,    Kong A. K.~H.,  2014, MNRAS, 437,
  1208

\bibitem[\protect\citeauthoryear{Farrell, Webb, Barret, Godet \&
  Rodrigues}{Farrell et~al.}{2009}]{farrell_intermediate-mass_2009}
Farrell S.~A.,  Webb N.~A.,  Barret D.,  Godet O.,    Rodrigues J.~M.,  2009,
  Nature, 460, 73

\bibitem[\protect\citeauthoryear{Godet, Barret, Webb, Farrell \& Gehrels}{Godet
  et~al.}{2009}]{godet_first_2009}
Godet O.,  Barret D.,  Webb N.~A.,  Farrell S.~A.,    Gehrels N.,  2009, ApJ
  Letters, 705, L109

\bibitem[\protect\citeauthoryear{Godet, Lombardi, Antonini, Barret, Webb,
  Vingless \& Thomas}{Godet et~al.}{2014}]{godet_implications_2014}
Godet O.,  Lombardi J.~C.,  Antonini F.,  Barret D.,  Webb N.~A.,  Vingless J.,
     Thomas M.,  2014, ApJ, 793, 105

\bibitem[\protect\citeauthoryear{Godet, Plazolles, Kawaguchi, Lasota, Barret,
  Farrell, Braito, Servillat, Webb \& Gehrels}{Godet
  et~al.}{2012}]{godet_investigating_2012}
Godet O.,  Plazolles B.,  Kawaguchi T.,  Lasota J.-P.,  Barret D.,  Farrell
  S.~A.,  Braito V.,  Servillat M.,  Webb N.,    Gehrels N.,  2012, ApJ, 752,
  34

\bibitem[\protect\citeauthoryear{Hut}{Hut}{1981}]{hut_tidal_1981}
Hut P.,  1981, A\&A, 99, 126

\bibitem[\protect\citeauthoryear{Ivanov \& Papaloizou}{Ivanov \&
  Papaloizou}{2004}]{ivanov_tidal_2004}
Ivanov P.~B.,  Papaloizou J. C.~B.,  2004, MNRAS, 347, 437

\bibitem[\protect\citeauthoryear{King \& Lasota}{King \&
  Lasota}{2014}]{king_hlx-1_2014}
King A.,  Lasota J.-P.,  2014, MNRAS, 444, L30

\bibitem[\protect\citeauthoryear{Kippenhahn, Weigert \& Weiss}{Kippenhahn
  et~al.}{2012}]{kippenhahn_stellar_2012}
Kippenhahn R.,  Weigert A.,    Weiss A.,  2012, Stellar {Structure} and
  {Evolution}

\bibitem[\protect\citeauthoryear{Kong, Soria \& Farrell}{Kong
  et~al.}{2015}]{kong_new_2015}
Kong A. K.~H.,  Soria R.,    Farrell S.~A.,  2015, The Astronomer's Telegram,
  6916, 1

\bibitem[\protect\citeauthoryear{Lajoie \& Sills}{Lajoie \&
  Sills}{2011}]{lajoie_mass_2011}
Lajoie C.-P.,  Sills A.,  2011, ApJ, 726, 67

\bibitem[\protect\citeauthoryear{Lasota, Alexander, Dubus, Barret, Farrell,
  Gehrels, Godet \& Webb}{Lasota et~al.}{2011}]{lasota_origin_2011}
Lasota J.-P.,  Alexander T.,  Dubus G.,  Barret D.,  Farrell S.~A.,  Gehrels
  N.,  Godet O.,    Webb N.~A.,  2011, ApJ, 735, 89

\bibitem[\protect\citeauthoryear{Lombardi, Sills, Rasio \& Shapiro}{Lombardi
  et~al.}{1999}]{lombardi_tests_1999}
Lombardi J.~C.,  Sills A.,  Rasio F.~A.,    Shapiro S.~L.,  1999, Journal of
  Computational Physics, 152, 687

\bibitem[\protect\citeauthoryear{MacLeod, Ramirez-Ruiz, Grady \&
  Guillochon}{MacLeod et~al.}{2013}]{macleod_spoon-feeding_2013}
MacLeod M.,  Ramirez-Ruiz E.,  Grady S.,    Guillochon J.,  2013, ApJ, 777, 133

\bibitem[\protect\citeauthoryear{Meibom \& Mathieu}{Meibom \&
  Mathieu}{2005}]{meibom_robust_2005}
Meibom S.,  Mathieu R.~D.,  2005, ApJ, 620, 970

\bibitem[\protect\citeauthoryear{Miller \& Colbert}{Miller \&
  Colbert}{2004}]{miller_intermediate-mass_2004}
Miller M.~C.,  Colbert E. J.~M.,  2004, International Journal of Modern Physics
  D, 13, 1

\bibitem[\protect\citeauthoryear{Miller, Farrell \& Maccarone}{Miller
  et~al.}{2014}]{miller_wind_2014}
Miller M.~C.,  Farrell S.~A.,    Maccarone T.~J.,  2014, ApJ, 788, 116

\bibitem[\protect\citeauthoryear{Monaghan}{Monaghan}{1992}]{monaghan_smoothed_1992}
Monaghan J.~J.,  1992, ARA\&A, 30, 543

\bibitem[\protect\citeauthoryear{Morris \& Monaghan}{Morris \&
  Monaghan}{1997}]{morris_switch_1997}
Morris J.~P.,  Monaghan J.~J.,  1997, Journal of Computational Physics, 136, 41

\bibitem[\protect\citeauthoryear{Paxton, Bildsten, Dotter, Herwig, Lesaffre \&
  Timmes}{Paxton et~al.}{2011}]{paxton_modules_2011}
Paxton B.,  Bildsten L.,  Dotter A.,  Herwig F.,  Lesaffre P.,    Timmes F.,
  2011, ApJS, 192, 3

\bibitem[\protect\citeauthoryear{Pelupessy}{Pelupessy}{2005}]{pelupessy_numerical_2005}
Pelupessy F.~I.,  2005, PhD thesis, Leiden Observatory, Leiden University, P.O.
  Box 9513, 2300 RA Leiden, The Netherlands

\bibitem[\protect\citeauthoryear{Pelupessy, van Elteren, de Vries, McMillan,
  Drost \& Portegies~Zwart}{Pelupessy
  et~al.}{2013}]{pelupessy_astrophysical_2013}
Pelupessy F.~I.,  van Elteren A.,  de Vries N.,  McMillan S. L.~W.,  Drost N.,
    Portegies~Zwart S.~F.,  2013, A\&A, 557, 84

\bibitem[\protect\citeauthoryear{Peters}{Peters}{1964}]{peters_gravitational_1964}
Peters P.-C.,  1964, Phys. Rev., 136, B1224

\bibitem[\protect\citeauthoryear{Portegies~Zwart, McMillan, van Elteren,
  Pelupessy \& de Vries}{Portegies~Zwart
  et~al.}{2013}]{portegies_zwart_multi-physics_2013}
Portegies~Zwart S.,  McMillan S. L.~W.,  van Elteren E.,  Pelupessy I.,    de
  Vries N.,  2013, Computer Physics Communications, 183, 456

\bibitem[\protect\citeauthoryear{Portegies~Zwart \& Verbunt}{Portegies~Zwart \&
  Verbunt}{2012}]{portegies_zwart_seba:_2012}
Portegies~Zwart S.~F.,  Verbunt F.,  2012, {SeBa}: {Stellar} and binary
  evolution

\bibitem[\protect\citeauthoryear{Price}{Price}{2012}]{price_smoothed_2012}
Price D.~J.,  2012, Journal of Computational Physics, 231, 759

\bibitem[\protect\citeauthoryear{Rasio, Tout, Lubow \& Livio}{Rasio
  et~al.}{1996}]{rasio_tidal_1996}
Rasio F.~A.,  Tout C.~A.,  Lubow S.~H.,    Livio M.,  1996, ApJ, 470, 1187

\bibitem[\protect\citeauthoryear{Regos, Bailey \& Mardling}{Regos
  et~al.}{2005}]{regos_mass_2005}
Regos E.,  Bailey V.~C.,    Mardling R.,  2005, MNRAS, 358, 544

\bibitem[\protect\citeauthoryear{Reimers}{Reimers}{1975}]{reimers_circumstellar_1975}
Reimers D.,  1975, Memoires of the Societe Royale des Sciences de Liege, 8, 369

\bibitem[\protect\citeauthoryear{Sepinsky, Willems \& Kalogera}{Sepinsky
  et~al.}{2007}]{sepinsky_equipotential_2007}
Sepinsky J.~F.,  Willems B.,    Kalogera V.,  2007, ApJ, 660, 1624

\bibitem[\protect\citeauthoryear{Sepinsky, Willems, Kalogera \& Rasio}{Sepinsky
  et~al.}{2007}]{sepinsky_interacting_2007}
Sepinsky J.~F.,  Willems B.,  Kalogera V.,    Rasio F.~A.,  2007, ApJ, 667,
  1170

\bibitem[\protect\citeauthoryear{Sepinsky, Willems, Kalogera \& Rasio}{Sepinsky
  et~al.}{2009}]{sepinsky_interacting_2009}
Sepinsky J.~F.,  Willems B.,  Kalogera V.,    Rasio F.~A.,  2009, ApJ, 702,
  1387

\bibitem[\protect\citeauthoryear{Servillat, Farrell, Lin, Godet, Barret \&
  Webb}{Servillat et~al.}{2011}]{servillat_x-ray_2011}
Servillat M.,  Farrell S.~A.,  Lin D.,  Godet O.,  Barret D.,    Webb N.~A.,
  2011, ApJ, 743, 6

\bibitem[\protect\citeauthoryear{Shakura \& Sunyaev}{Shakura \&
  Sunyaev}{1973}]{shakura_black_1973}
Shakura N.~I.,  Sunyaev R.~A.,  1973, A\&A, 24, 337

\bibitem[\protect\citeauthoryear{Soria}{Soria}{2013}]{soria_eccentricity_2013}
Soria R.,  2013, MNRAS, 428, 1944

\bibitem[\protect\citeauthoryear{Soria, Hau \& Pakull}{Soria
  et~al.}{2013}]{soria_kinematics_2013}
Soria R.,  Hau G. K.~T.,    Pakull M.~W.,  2013, ApJ Letters, 768, L22

\bibitem[\protect\citeauthoryear{Straub, Godet, Webb, Servillat \&
  Barret}{Straub et~al.}{2014}]{straub_investigating_2014}
Straub O.,  Godet O.,  Webb N.,  Servillat M.,    Barret D.,  2014, A\&A, 569,
  116

\bibitem[\protect\citeauthoryear{van Elteren, Pelupessy \& Portegies~Zwart}{van
  Elteren et~al.}{2014}]{van_elteren_multi-scale_2014}
van Elteren A.,  Pelupessy I.,    Portegies~Zwart S.,  2014, Royal Society of
  London Philosophical Transactions Series A, 372, 30385

\bibitem[\protect\citeauthoryear{Webb, Cseh, Lenc, Godet, Barret, Corbel,
  Farrell, Fender, Gehrels \& Heywood}{Webb et~al.}{2012}]{webb_radio_2012}
Webb N.,  Cseh D.,  Lenc E.,  Godet O.,  Barret D.,  Corbel S.,  Farrell S.,
  Fender R.,  Gehrels N.,    Heywood I.,  2012, Science, 337, 554

\bibitem[\protect\citeauthoryear{Webb, Barret, Godet, Servillat, Farrell \&
  Oates}{Webb et~al.}{2010}]{webb_chandra_2010}
Webb N.~A.,  Barret D.,  Godet O.,  Servillat M.,  Farrell S.~A.,    Oates
  S.~R.,  2010, ApJ Letters, 712, L107

\bibitem[\protect\citeauthoryear{Wiersema, Farrell, Webb, Servillat, Maccarone,
  Barret \& Godet}{Wiersema et~al.}{2010}]{wiersema_redshift_2010}
Wiersema K.,  Farrell S.~A.,  Webb N.~A.,  Servillat M.,  Maccarone T.~J.,
  Barret D.,    Godet O.,  2010, ApJ Letters, 721, L102

\bibitem[\protect\citeauthoryear{Zahn}{Zahn}{1977}]{zahn_tidal_1977}
Zahn J.-P.,  1977, A\&A, 57, 383

\end{thebibliography}
\bibliographystyle{mn2e}

\end{document}